\documentclass[11pt,a4paper]{article}
\pdfoutput=1

\usepackage{jcappub}
\usepackage{amsmath,amssymb,color,bm, graphicx}
\usepackage{enumerate,xstring, hyperref}

\def\be{\begin{equation}}
\def\ee{\end{equation}}
\def\bea{\begin{eqnarray}}
\def\eea{\end{eqnarray}}

\def\ba#1\ea{\begin{align}#1\end{align}}
\def\bg#1\eg{\begin{gather}#1\end{gather}}

\renewcommand{\v}[1]{\bm{#1}}

\newcommand{\vx}{\v{x}}
\newcommand{\vv}{\v{v}}

\newcommand{\vk}{\v{k}}
\newcommand{\vq}{\v{q}}

\renewcommand{\d}{\delta}

\newcommand{\fidu}{Fiducial}
\newcommand{\high}{High}
\newcommand{\loww}{Low}

\def\be{\begin{equation}}
\def\ee{\end{equation}}
\def\ben{\begin{eqnarray}}
\def\een{\end{eqnarray}}
\def\ba{\begin{array}}
\def\ea{\end{array}}

\def\ba#1\ea{\begin{align}#1\end{align}}

\newcommand{\bq}{\begin{eqnarray}}
\newcommand{\eq}{\end{eqnarray}}
\newcommand{\bes}{\begin{subequations}}
\newcommand{\ees}{\end{subequations}}

\def\R{\mathcal{R}}

\def\O{\mathcal{O}}

\makeatletter
\newlength{\apb@width}
\newcommand{\autoparbox}[2][c]{\settowidth{\apb@width}{#2}\parbox[#1]{\apb@width}{#2}}

\makeatother

\newcommand{\comment}[1]{}

\definecolor{seagreen}{rgb}{0.18, 0.55, 0.34}

\begin{document}

\title{Baryon-CDM isocurvature galaxy bias with IllustrisTNG}

\author{Alexandre Barreira,$^1$}
\emailAdd{barreira@MPA-Garching.MPG.DE}
\affiliation{$^1$Max-Planck-Institut f{\"u}r Astrophysik, Karl-Schwarzschild-Str.~1, 85741 Garching, Germany}

\author{Giovanni Cabass,$^1$}
\emailAdd{gcabass@MPA-Garching.MPG.DE}

\author{Dylan Nelson,$^1$}
\emailAdd{dnelson@MPA-Garching.MPG.DE}

\author{Fabian~Schmidt$^1$}
\emailAdd{fabians@MPA-Garching.MPG.DE}

\abstract{We study the impact that baryon-CDM relative density perturbations $\delta_{bc}$ have on galaxy formation using cosmological simulations with the IllustrisTNG model. These isocurvature (non-adiabatic) perturbations can be induced primordially, if multiple fields are present during inflation, and are generated before baryon-photon decoupling when baryons did not comove with CDM. The presence of long-wavelength $\delta_{bc}$ perturbations in our simulations is mimicked by modifying the ratios of the cosmic densities of baryons $\Omega_b$ and CDM $\Omega_c$, at fixed total matter density $\Omega_m$. We measure the corresponding galaxy bias parameter $b_{\delta}^{bc}$ as the {\it response} of galaxy abundances to $\delta_{bc}$. When selecting by total host halo mass, $b_{\delta}^{bc}$ is negative and it decreases with mass and redshift. Stellar-mass selected simulated galaxies show a weaker or even the opposite trend because of the competing effects of $\delta_{bc}$ on the halo mass function and stellar-to-halo-mass relations. We show that simple modeling of the latter two effects describes $b_{\delta}^{bc}$ for stellar-mass-selected objects well. We find $b_{\delta}^{bc} =0.6$ for $M_* = 10^{11}\ M_{\odot}/h$ and $z=0.5$, which is representative of BOSS DR12 galaxies. For $\delta_{bc}$ modes generated by baryon-photon interactions, we estimate the impact on the DR12 power spectrum to be below $1\%$, and shifts on inferred distance and growth rate parameters should not exceed $0.1\%$.}


\date{\today}

\maketitle
\flushbottom


\section{Introduction}\label{sec:intro}

The study of the large-scale clustering of galaxies in the Universe is one of the most promising avenues to address long-standing open issues in fundamental physics such as the nature of dark matter and dark energy, what is the law of gravity on large scales and what is the value of neutrino masses. One of the main ingredients in the exercise of theoretically predicting the statistics of the galaxy distribution is a description of {\it galaxy bias}, i.e.~a formalism that relates the observed distribution of galaxies to that of the underlying matter field (see Ref.~\cite{biasreview} for a comprehensive review). Formally, the density contrast of the galaxies at position $\vx$ and redshift $z$, $\delta_g(\vx, z)$, can be expanded as
\bq\label{eq:bias_exp_intro}
\delta_g(\vx, z) = \sum_{\O} b_\O(z) \O(\vx, z),
\eq
where the sum runs with all generality over all types of large-scale perturbations $\O(\vx, z)$ that can have an impact on galaxy formation. This expansion should also include stochastic terms, as well as projection and selection effects \cite{2018JCAP...12..035D}, but we refrain from writing these explicitly here. Physically, the bias parameters\footnote{The word ``parameter'' is a misnomer since they depend on time and also on galaxy properties (such as mass, luminosity, etc.). We nonetheless retain this nomenclature as it is usually adopted in the literature.} encode all of the dependence of galaxy formation processes on the large-scale perturbations that each of them multiplies; more technically, the $b_\O(z)$ specify how the galaxy distribution ``responds'' to changes in the amplitude of the perturbations $\O(\vx, z)$. 

A central question prior to any galaxy clustering study concerns therefore the number of terms that should be taken into account in the expansion of Eq.~(\ref{eq:bias_exp_intro}). For example, the so-called {\it local-in-matter-density} (LIMD) bias parameters correspond to including terms such as \cite{fry/gaztanaga:1983} $\delta_g(\vx, z) \supset \sum_{n} b_n(z) \delta^n_m(\vx, z)/n!$, where $\delta_m(\vx, z) \propto \nabla^2\Phi(\vx, z)$ is the total matter density contrast ($\Phi$ is the gravitational potential). This captures all of the dependency of galaxy formation on the amplitude of large-scale isotropic perturbations. The order $n$ up to which one should keep terms is determined by the order in perturbation theory \cite{Bernardeau/etal:2002} up to which one wishes to model the statistics of the galaxy distribution, which is in turn partly determined by how deep into the nonlinear regime of structure formation one wishes to analyse some given galaxy sample. For instance, to model the galaxy power spectrum at leading order (or tree-level) in perturbation theory, one would need to retain only the first-order term $\propto \delta_m$; next-to-leading order (or $1$-loop), one would need to include also the term $\propto \delta_m^2$, etc. In addition to the LIMD bias parameters, there are also important contributions from large-scale tidal fields 
\cite{mcdonald/roy:2009, chan/scoccimarro/sheth:2012, baldauf/etal:2012, saito/etal:14}, e.g.~$\delta_g(\vx, z) \supset b_{K^2}K^2_{ij}(\vx, z)$, where $K_{ij}(\vx, z) = (\partial_i\partial_j/\nabla^2 - \delta_{ij}/3)\delta_m$; 
as well as from $\O(\vx, z)$ operators constructed from higher than second order derivatives of the potential such as $\nabla^2\delta_m(\vx, z)$ \cite{2019arXiv190411294L}. The bottom line is that, at a given order in perturbation theory, it is important to make sure that one enumerates all of the possible operators $\O(\vx, z)$ in Eq.~(\ref{eq:bias_exp_intro}) that can influence galaxy formation.

All the bias parameters mentioned so far have in common the fact that they are associated with perturbations that depend on the gravitational potential that is sourced by total matter density fluctuations (i.e., growing adiabatic perturbations); these are also the most widely studied bias parameters in the literature. In this paper, we turn our attention instead to relative perturbations between the baryon and cold dark matter (CDM) components, which have received less attention in the literature despite being potentially relevant. {These isocurvature perturbations can be generated during inflation \cite{grin/dore/kamionkowski} in multifield scenarios \cite{1997PhRvD..56..535L, 2000PhRvD..62d3504L, 2003PhRvD..67b3503L, 2006RvMP...78..537B}. In addition, even for purely adiabatic primordial perturbations, they are sourced before baryon-photon decoupling, when baryons did not move along the same trajectories as dark matter because of their tight coupling to the photons. For both generation mechanisms, the consequences are} long-wavelength modulations of the amount of baryons relative to CDM after decoupling \cite{barkana/loeb:11, 2016PhRvD..94f3508S, 2016ApJ...830...68A, 2019JCAP...06..006C}. Given that galaxy formation and evolution is sensitive to the amount of baryons available to participate in processes such as baryonic accretion, star formation, black hole growth, and feedback, it is therefore important to study the impact of these {\it baryon-CDM density perturbations}. Likewise, after decoupling, there will also be regions in the Universe that exhibit relative {\it baryon-CDM velocity perturbations}, i.e.~patches within which baryons and CDM move at different velocities \cite{tseliakhovich/hirata:2010, blazek/etal:15, 2016PhRvD..94f3508S}. A noteworthy aspect of baryon-CDM perturbations is that they exhibit strong baryon acoustic oscillation (BAO) features that are not completely in phase with those imprinted in the total matter fluctuations. This can lead to shifts in the BAO scale imprinted in the galaxy distribution, which should be accounted for to guard against potential biases on cosmological parameters. 

More concretely, our main objective in this paper is to estimate the baryon-CDM density perturbation bias parameter $b_{\delta}^{bc}$. We shall do so by carrying out hydrodynamical cosmological simulations of galaxy formation with the {\sc AREPO} code \citep{2010MNRAS.401..791S, 2016MNRAS.455.1134P} and the IllustrisTNG physics model \citep{2017MNRAS.465.3291W, Pillepich:2017jle} in the presence of long-wavelength baryon-CDM density perturbations. These perturbations can be incorporated in the simulations by perturbing, relative to some fiducial cosmology, the cosmic fractions of baryons $\Omega_b$ and CDM $\Omega_c$, while keeping the total matter density $\Omega_m$ fixed. By way of the separate universe formalism, galaxy formation and evolution in this modified cosmology is equivalent to that taking place inside a long-wavelength baryon-CDM density perturbation in the fiducial cosmology. We will see that the effect of a baryon-CDM density perturbation impacts galaxy formation in two main ways: (i) the modified baryon-to-CDM ratio alters the shape of the initial matter power spectrum; and (ii) a modified baryon density alters the {\it fuel} supply for star formation, which results in modified stellar-to-halo-mass relations, as well as modified onset times for baryon feedback processes (such as active galactic nuclei (AGN) or supernovae feedback).

The amplitude of the baryon-CDM density bias parameter $b_{\delta}^{bc}$ has been previously estimated with simple analytical arguments in Refs.~\cite{barkana/loeb:11, 2016PhRvD..94f3508S}. The results that we present in this paper are, to the best of our knowledge, the first ever measurement of the baryon-CDM density bias from state-of-the-art galaxy formation simulations. Our numerical results will show that $b_{\delta}^{bc}$ can be sizeable for galaxy/halo mass scales and redshifts that are relevant to both current and future galaxy redshift surveys (while being well within the observational bounds reported by  Ref.~\cite{2017MNRAS.470.2723B}; see also Refs.~\cite{soumagnac/etal:16, 2019MNRAS.485.1248S}). This, together with the fact that $b_{\delta}^{bc}$ enters the bias expansion of Eq.~(\ref{eq:bias_exp_intro}) at leading order, makes it important to investigate the impact of including baryon-CDM density perturbations in models of galaxy clustering \cite{2016ApJ...830...68A, 2017MNRAS.470.2723B, 2019JCAP...06..006C}. {For adiabatic initial conditions after inflation,} our results will show, however, that the impact of $\delta_{bc}$ on the galaxy power spectrum is not expected to exceed the $1\%$ level for mass scales and redshifts relevant for current and future galaxy surveys. These $\delta_{bc}$ perturbations are also not expected to bias distance, Hubble rate and growth rate measurements from galaxy samples like BOSS DR12 by more than $0.1\%$. Here, we do not study the impact of baryon-CDM velocity perturbations and we defer such an investigation to future work (see, however, Refs.~\cite{tseliakhovich/hirata:2010, yoo/etal:2011, yoo/seljak, tseliakhovich/barkana/hirata, dalal/etal:2010, slepian/eisenstein, asaba/ichiki/tashiro, blazek/etal:15, Slepian:2016nfb, 2016PhRvD..94f3508S, 2019JCAP...06..006C} for such past studies, including Refs.~\cite{popa/etal, 2011ApJ...736..147G, 2011MNRAS.412L..40M, 2011ApJ...730L...1S, 2012ApJ...747..128N, Visbal/etal:12, 2012ApJ...760....4O, 2013ApJ...763...27N, 2013ApJ...771...81R, 2019ApJ...878L..23C} for investigations with $N$-body simulations and Ref.~\cite{Fialkov:2014rba} for a review).

This paper is organized as follows. In Sec.~\ref{sec:sims}, we outline the main aspects of how baryon-CDM perturbations can impact large-scale structure formation and describe the numerical simulation setup we adopt to determine its associated density bias parameter $b_{\delta}^{bc}$. Section~\ref{sec:res} contains our main numerical results: in Sec.~\ref{sec:res:totmass}, we present and discuss our results using simulated galaxies selected by their total host halo mass, whereas in Sec.~\ref{sec:res:stemass}, we do the same but selecting galaxies by their stellar mass. In Sec.~\ref{sec:Pgk}, we estimate the impact that baryon-CDM density perturbations can have on the galaxy power spectrum. We summarize and conclude in Sec.~\ref{sec:conc}. In App.~\ref{app:baryon-CDM-theory}, we provide more details about the equations of baryon-CDM perturbations and in App.~\ref{app:stecorrection}, we describe a resolution correction strategy that we implement to self-consistently compare galaxy stellar mass values at different IllustrisTNG resolutions. 


\section{Baryon-CDM perturbations and Separate Universe simulations}\label{sec:sims}

In this section, we lay down the basics of the contribution of baryon-CDM perturbations to the general galaxy bias expansion and anticipate the main physical impact they will have on galaxy formation. We also describe the separate universe simulations that we perform with the IllustrisTNG model to obtain our numerical results. 


\subsection{Baryon-CDM perturbations in the galaxy bias expansion}\label{sec:sims:theory}

Prior to the epoch of recombination (or more precisely, baryon-photon decoupling), the baryon and CDM components behave differently: the non-interacting CDM component is collisionless and experiences only gravity, but the baryons are tightly coupled to the photons and experience additional pressure forces that keep them from following the same trajectories as CDM. In other words, after baryons decouple from the photons, they are not comoving with the CDM component and cannot therefore be strictly treated as a single fluid \cite{shoji/komatsu, somogyi/smith:2010, bernardeau/vdr/vernizzi, lewandowski/perko/senatore}. A practical consequence of this that is relevant for galaxy formation is that there will be regions in the Universe that exhibit relative velocity \cite{tseliakhovich/hirata:2010, blazek/etal:15, 2016PhRvD..94f3508S} and relative density perturbations \cite{barkana/loeb:11, 2016PhRvD..94f3508S, 2016ApJ...830...68A} between the baryon and CDM components. These perturbations, which we call {\it baryon-CDM perturbations} here, are not normally taken into account in studies of large-scale structure formation, but it is important to move beyond (or at least assess the degree of validity of) this approximation since galaxy formation physics depends sensitively on the relative amounts of baryons and CDM. {These perturbations are guaranteed to arise because of the baryon-photon interactions, but relative density perturbations can also be produced earlier if multiple fields are present during inflation; in the literature, these are often called compensated isocurvature perturbations (CIP) \cite{2009PhRvD..80f3535G, 2010ApJ...716..907H, grin/dore/kamionkowski, 2014PhRvD..89b3006G, 2017PhRvD..96h3508S, 2016PhRvD..93d3008M, 2019MNRAS.485.1248S, 2019arXiv190400024H}.}

In terms of the bias expansion of Eq.~(\ref{eq:bias_exp_intro}), these baryon-CDM perturbations contribute to first order as
\bq\label{eq:bias_exp1}
\delta_g(\vx, z) \supset b_\delta^{bc}(z) \delta_{bc}(\vx) + b_\theta^{bc}(z) \theta_{bc}(\vx, z),
\eq
where $\delta_{bc}(\vx)$ is a constant compensated baryon-CDM perturbation characterized by $\delta_c = -f_b\delta_b$, $\delta_m = 0$ with $\delta_b$, $\delta_c$ the density contrasts of baryons and CDM, respectively, and $f_b = \Omega_b/\Omega_c$ (we neglect neutrino masses, so that neutrinos do not contribute to matter). The term $\theta_{bc}(\vx, z)$,  on the other hand, corresponds to a relative velocity divergence $\theta_{bc} = {\bm\nabla}\cdot\vv_{bc}$, with $\vv_{bc} = \vv_b - \vv_c$ the relative velocity between baryons and CDM; the amplitude of this mode decays with time (see App.~\ref{app:baryon-CDM-theory} for more details). There is already a significant literature \cite{barkana/loeb:11, blazek/etal:15, 2016PhRvD..94f3508S, 2016ApJ...830...68A} discussing these terms, as well as some observational constraints. For example, Ref.~\cite{2017MNRAS.470.2723B} using the galaxy power spectrum from the BOSS DR12 sample finds {(assuming photon-baryon interactions as the production mechanism)} $b_{\delta}^{bc} = -1.0 \pm 6.2$ and $b_{\theta}^{bc} = -114 \pm 175$ at the $95\%$ confidence level. In Refs.~\cite{soumagnac/etal:16, 2019MNRAS.485.1248S} the authors looked for the impact of baryon-CDM density perturbations by comparing number- and luminosity-weighted galaxy statistics, which are expected to be sensitive to baryon-CDM perturbations; the analysis of Ref.~\cite{2019MNRAS.485.1248S} is also consistent with a null detection. 

In this paper, we focus on the bias parameter $b_{\delta}^{bc}$, i.e.~we are interested in studying galaxy formation within constant baryon-CDM density perturbations. According to the estimate of Ref.~\cite{2016PhRvD..94f3508S}, this is also the term that is expected to be the most important in the bias expansion {(see also Ref.~\cite{2019JCAP...06..006C})}. We will further be interested in cases in which the wavelength of these perturbations is assumed to be sufficiently large compared to the typical scale of galaxy formation.\footnote{If $R_*$ denotes the size of this scale, then the requirement is for the wavenumber of the baryon-CDM perturbations to be sufficiently smaller than $2\pi/R_*$. For halos, this scale is of order the size of their Lagrangian radius, $R_* = R_{\rm Lag.}^{\rm halo} \sim 10\ {\rm Mpc}/h$, which implies $k \ll 2\pi/R_* \approx 0.6\ h/{\rm Mpc}$. The same scale for galaxies is however more uncertain and not necessarily the same. For instance, Ref.~\cite{2019JCAP...05..031C} noted that radiative-transfer effects during reionization introduce a new scale in the bias expansion for galaxies, of order the mean free path of ionizing photons.} ``Sufficiently'' here means that the modulation of the baryon-CDM perturbations effectively acts as a modified background to the physical processes that govern galaxy formation. In this limit, one can then make use of the separate universe ansatz to gain intuition about the expected phenomenology, as well as to setup the numerical simulations needed to study the effect (see the next subsection). The separate universe ansatz states that {\it local structure formation inside a long-wavelength perturbation in some fiducial cosmology is equivalent to global structure formation in an appropriately modified cosmology}. For the case of baryon-CDM density perturbations, the modifications to the cosmology involve altering the background densities of baryons and CDM, $\Omega_b$, $\Omega_c$, while keeping the total matter density $\Omega_m = \Omega_b + \Omega_c$ the same. Physically, one can then anticipate two main ways through which baryon-CDM density perturbations impact large-scale structure formation:

\begin{enumerate}
\item[(1)] Changes in the relative sizes of $\Omega_b$, $\Omega_c$ lead to a difference in shape of the linear matter power spectrum on scales $k \gtrsim k_{\rm eq} \approx 0.02\ h/{\rm Mpc}$ (larger $\Omega_{b}$ suppresses small-scale power).  This effect impacts structure formation, even if the latter takes place in a purely gravitational, pressureless manner. 

\item[(2)] Different baryonic densities $\Omega_b$ will also modify the fraction of the total amount of non-relativistic matter that can experience non-gravitational forces, undergo radiative cooling, form stars and black holes, which in turn feedback onto the rest of the matter via supernovae explosions and gas ejected by AGN. 
\end{enumerate}

Effect (1) above can be split further into two main physical effects. First, at fixed total matter $\Omega_m$, an increase in the baryon density leads to a corresponding reduction in the dark matter density, and correspondingly less growth of structure between the end of radiation domination and baryon-photon decoupling. This suppresses the amplitude of the power spectrum on scales smaller than the horizon at the epoch of matter-radiation equality, $k > k_{\rm eq}$ (note that $k_{\rm eq}$ remains the same since $\Omega_m$ is fixed). A second physical effect is associated with the change in the sound speed of the photon-baryon plasma, which depends on the ratio of baryon to photon densities, and which impacts the total matter power spectrum via a modified BAO feature. This second effect has been studied recently in Ref.~\cite{2019arXiv190400024H} in the context of baryon-CDM density perturbations generated during inflation, i.e. primordial CIPs.

{For $\delta_{bc}$ perturbations generated during inflation and that are still outside the sound horizon at photon-baryon decoupling (cf.~App.~\ref{app:baryon-CDM-theory}), the size of effect (1) can be calculated directly using Einstein-Boltzmann codes with modified baryon and CDM cosmic fractions; this is the approach we take in this paper. For perturbations that are inside the horizon at the time of decoupling, including those generated by baryon-photon interactions, the calculation of the initial power spectrum is more involved, as $\d_{bc}$ then evolves with time in a scale-dependent way before converging to a constant $\delta_{bc}$ value after recombination is complete.  This means that the effect of a large-scale $\d_{bc}$ mode on the evolution of small-scale modes before decoupling, which underlies the effect (1), depends on the wavelength of the mode and cannot be captured precisely by varying $\Omega_b/\Omega_c$; the latter can only describe the regime where $\d_{bc}$ is constant in time. 
  We argue in App.~\ref{app:baryon-CDM-theory} that this fact
  actually leads to an overestimate of the impact of effect (1) above for $\delta_{bc}$ modes generated solely due to baryon-photon interactions. Effect (2), on the other hand, does not depend on the exact past evolution of $\delta_{bc}$, but only on its value at the starting time of the simulation. We will return to these points whenever relevant to the interpretation of our results below.}


\subsection{Separate universe simulations of baryon-CDM perturbations}\label{sec:sims:sims}

\begin{table}
\centering
\begin{tabular}{@{}lccccccccccc}
\hline\hline
\rule{0pt}{1\normalbaselineskip}
Name &\ \ $\Omega_{m}$ & \ \ $\Omega_{b}$ & \ \ $\Omega_{c}$ & \ \ $\Omega_{\Lambda}$ & \ \ $h$ & \ \ $n_s$ & \ \ $A_s$
\\
\hline
\rule{0pt}{1\normalbaselineskip}
\fidu &\ \ $0.3089$ & \ \ $0.0486$ & \ \ $0.2603$ & \ \ $0.6911$ & \ \ $0.6774$ & \ \ $0.967$ & \ \ $2.068 \times 10^{-9}$ 
\\
\\
\,\,\high &\ \ " & \ \ $0.0510$ & \ \ $0.2579$ & \ \ " & \ \ " & \ \ " & \ \ " 
\\
\\
\,\,\loww &\ \ " & \ \ $0.0462$ & \ \ $0.2627$ & \ \ " & \ \ " & \ \ " & \ \ " 
\\
\hline
\hline
\end{tabular}
\caption{Parameters of the cosmologies simulated in this paper. The \high\ and \loww\ cosmologies describe the effect of $\Delta_b = 0.05$ and $\Delta_b = -0.05$ long-wavelength baryon-CDM density perturbations in the \fidu\ cosmology (cf.~Eqs.~(\ref{eq:sepuni_para}) and (\ref{eq:relpara})). Note that all of the cosmological parameters are the same, except $\Omega_b$ and $\Omega_c$ (this is the meaning of "). We have simulated structure formation in these cosmologies at two particle resolutions: $N_p = 1250^3$, $L_{\rm box} = 75\ {\rm Mpc}/h$ (called TNG100-1.5) and $N_p = 1250^3$, $L_{\rm box} = 205\ {\rm Mpc}/h$ (called TNG300-2). Each simulation was also run without (dubbed Gravity) and with (dubbed Hydro) hydrodynamical physical processes (note that for the Hydro runs, the number of mass elements is twice the quoted values: $N_p$ gas cells and $N_p$ dark matter mass elements). The same random seed was used to generate the initial conditions of all the simulations. The value of the primordial power spectrum amplitude $A_s$ is evaluated at a pivot scale $k_{\rm pivot} = 0.05\ {\rm Mpc}^{-1}$; this value yields $\sigma_8(z=0) = 0.816$ in the \fidu\ cosmology.}
\label{table:params}
\end{table}

As already mentioned above, the effects of baryon-CDM density perturbations on galaxy formation can be mimicked by changes in the $\Omega_b$ and $\Omega_c$ cosmological parameters.\footnote{The implementation of the separate universe technique for the case of baryon-CDM density perturbations is even more straightforward than for the case of total density perturbations \cite{li/hu/takada, 2014PhRvD..90j3530L, wagner/etal:2014, CFCpaper2, baldauf/etal:2015, response, lazeyras/etal, li/hu/takada:2016, 2018PhRvD..97l3526C, 2019arXiv190402070B}. In the latter, the changes in cosmology alter the relation between comoving and physical distances, as well as the relation between redshift and physical time, which implies performing additional (although straightforward) adjustments to the simulation box size, desired output redshift values and structure-finding criteria. These are steps that we do not have to worry about here as baryon-CDM density perturbations keep the value of $\Omega_m$ unchanged.} Here, we describe these changes with a parameter $\Delta_b$ as
\bq\label{eq:sepuni_para}
\tilde{\Omega}_b &=& \Omega_b \left[1 + \Delta_b\right], \nonumber \\
\tilde{\Omega}_c &=& \Omega_c \left[1 - f_b \Delta_b\right],
\eq
where a tilde indicates a quantity in the modified cosmology. The value of $\Delta_b$ is related to the $\delta_{bc}$ term that enters the bias expansion Eq.~(\ref{eq:bias_exp1}) as
\bq\label{eq:relpara}
1 + \delta_{bc} &=& \frac{1 + \Delta_b}{1 - f_b\Delta_b} \approx 1 + \left(1 + f_b\right) \Delta_b \nonumber \\
&\Longrightarrow& \delta_{bc} = \left(1 + f_b\right) \Delta_b,
\eq
which follows from $\tilde{\Omega}_b / \tilde{\Omega}_c = [1+\delta_{bc}]\Omega_b/\Omega_c$. The equations of motion show that the relative density perturbation is a constant mode (cf.~App.~\ref{app:baryon-CDM-theory}). Physically, this is because the ratio of baryon to CDM densities is conserved under gravitational evolution. It is thus consistently absorbed in modified cosmological parameters $\Omega_b$ and $\Omega_c$. Then, given the number density $n_g$ of galaxies selected according to some property (e.g.~total mass or stellar mass), the bias coefficient $b_{\delta}^{bc}$ can be computed as
\begin{equation}
\label{eq:bias_as_response}
b_{\delta}^{bc}(z) = \frac{1}{n_g(z)}\frac{\partial n_g(z)}{\partial\delta_{bc}}\bigg|_{\delta_{bc}=0}.
\end{equation}

Table \ref{table:params} summarizes the three cosmological scenarios we consider in this paper. The two separate universe cosmologies, which we call \high\ and Low, are obtained from the \fidu\ cosmology by considering $\Delta_b = \Delta_b^{\rm \high} = 0.05$ and $\Delta_b = \Delta_b^{\rm \loww}= -0.05$, respectively. These numerical values of $\Delta_b$ are chosen from a compromise between having sizeable effects in the simulations (i.e., sufficiently high signal-to-noise in the evaluation of Eq.~(\ref{eq:bias_as_response}); see also Eqs.~(\ref{eq:bcb_measure2}) and (\ref{eq:bcb_measure3}) below), while keeping higher-order corrections small. All cosmologies share the same numerical values of the matter and cosmological constant densities $\Omega_m, \Omega_\Lambda$, dimensionless present-day Hubble rate $h$, spectral index $n_s$ and amplitude of the primordial power spectrum $A_s$ (evaluated at a pivot scale $k_{\rm pivot} = 0.05\ {\rm Mpc}^{-1}$); the numerical value of $A_s$ is that which yields $\sigma_8(z=0) = 0.816$ in the \fidu\ cosmology.

We obtain our numerical results using the moving-mesh code {\sc AREPO} \citep{2010MNRAS.401..791S, 2016MNRAS.455.1134P} together with the IllustrisTNG galaxy formation model \citep{2017MNRAS.465.3291W, Pillepich:2017jle}, which is an improved version of its precursor Illustris \citep{2014MNRAS.445..175G, 2014MNRAS.444.1518V}. We refer the interested reader to Refs.~\cite{2018MNRAS.480.5113M, Pillepich:2017fcc, 2018MNRAS.477.1206N, 2018MNRAS.475..676S, Nelson:2017cxy} for the first results from the IllustrisTNG simulations, as well as Ref.~\cite{2019arXiv190402070B} for separate universe simulations of total matter density perturbations; see also Ref.~\cite{Nelson:2018uso} for an overview of the publicly available simulation data. 

We perform simulations at two mass resolutions: $N_p = 1250^3$ particles on a cubic box with $L_{\rm box} = 75\ {\rm Mpc}/h$ on a side, called TNG100-1.5 throughout, and $N_p = 1250^3$, $L_{\rm box} = 205\ {\rm Mpc/h}$, which we call TNG300-2. This resolution classification is similar to that adopted in the original IllustrisTNG runs; our TNG100-1.5 case falls in between the TNG100-1 ($N_p = 1820^3$) and TNG100-2 ($N_p = 910^3$) resolutions. For each resolution, we run gravity-only simulations (dubbed Gravity throughout), as well as hydrodynamical IllustrisTNG simulations (dubbed Hydro; for these, the number of "particles" is doubled: $N_p$ gas cells and $N_p$ dark matter particles). The Gravity runs are sensitive to effect (1) mentioned in Sec.~\ref{sec:sims:theory} in isolation, whereas the Hydro runs are sensitive to both effects (1) and (2). Comparing the two will thus indicate the relative importance of the two effects. 

We generate the initial conditions at the same initial redshift $z_i = 127$ for the \fidu, \high\ and \loww\ cosmologies with the {\sc N-GenIC} code \citep{2015ascl.soft02003S} using the Zel'dovich approximation. We use the {\sc CAMB} code \citep{camb} to compute the linear matter power spectra at $z=0$, which we scale back to $z_i$ assuming no cosmic radiation density and then give to {\sc N-GenIC} as input. We also use the same random white-noise seed in {\sc N-GenIC} to generate the initial conditions for the three cosmologies for each resolution. {Note that we run CAMB with constant $\Omega_{b}$ and $\Omega_{c}$, which as discussed in the previous subsection, amounts to considering the impact on the initial power spectrum of baryon-CDM perturbations generated during inflation (CIPs) and that are still outside the sound horizon at decoupling.}

There are two other points worth noting about our numerical methodology, which are important for the interpretation of our results below. One is that the gas and CDM mass elements at the starting redshift are initialized with the same density perturbations and velocities. That is, structure formation in our simulations begins with the baryons comoving with the CDM component (this is as in the original IllustrisTNG simulations; see, however, Refs.~\cite{2003MNRAS.344..481Y, 2013MNRAS.434.1756A, 2017MNRAS.467.4401V} for studies about the impact of initializing baryons and CDM with different transfer functions). The other point is that we do not modify any of the parameters of the IllustrisTNG physics model when we modify $\Omega_b$ and $\Omega_c$. Our separate universe simulations thus follow structure formation in the fiducial cosmology with the hydrodynamical processes as specified by the IllustrisTNG model, inside a region at cosmic mean total matter density where baryons and CDM are comoving, and whose abundances are not the cosmic mean ones. An interesting question that we leave to be addressed in future work concerns the dependence of the predicted galaxy bias values on the baryon physics implementation in hydrodynamical simulations of galaxy formation.

The galaxies in our simulations are identified as gravitationally bound groups of stars as determined by the {\sc SUBFIND} algorithm \cite{2001MNRAS.328..726S} in data snapshots produced at redshifts $z = 3, 2, 1, 0.5, 0$. Following Eq.~\eqref{eq:bias_as_response}, we measure the baryon-CDM density bias $b_{\delta}^{bc}$ 
as
\bq\label{eq:bcb_measure}
b_{\delta}^{bc}(z, M) = \frac{b_{\delta}^{bc, \rm \high}(z, M) + b_{\delta}^{bc, \rm \loww}(z, M)}{2},
\eq
with
\begin{align}
b_{\delta}^{bc, \rm \high}(z, M) &= \frac{1}{\delta_{bc} ^{\rm \high}}\left[\frac{N^{\rm \high}(z,M)}{N^{\rm \fidu}(z,M)} - 1\right], \label{eq:bcb_measure2}\\
b_{\delta}^{bc, \rm \loww}(z, M) &= \frac{1}{\delta_{bc} ^{\rm \loww}}\left[\frac{N^{\rm \loww}(z,M)}{N^{\rm \fidu}(z,M)} - 1\right], \label{eq:bcb_measure3}
\end{align}
where, recall, $\delta_{bc} = (1 + f_b)\Delta_b$ and $N(z,M)$ denotes the number of galaxies found in the corresponding cosmology at redshift $z$ in some mass bin centered at $M$. For each cosmology and resolution, we have only simulated a single realization of the initial conditions, which prevents us from quoting errors in a statistical-ensemble sense. Theoretically, the values of $b_{\delta}^{bc, \rm \high}$ and $b_{\delta}^{bc, \rm \loww}$ should be the same, and hence we shall use their difference as a rough guide for the error in our measurements. Note also that Eq.~(\ref{eq:bcb_measure}) corresponds to a central finite difference, and hence the numerical error is of order $\delta_{bc}^2$ ($\approx 0.35\%$ for $\Delta_b = 0.05$). 


\section{Numerical results: baryon-CDM density galaxy bias}\label{sec:res}

In this section, we show measurements of $b_{\delta}^{bc}$ for galaxies selected as a function of the mass of all particles that are gravitationally bound to the halo, $M_{\rm h}$, as well as a function of the total mass in stars enclosed within twice the stellar half-mass radius (the radius that encloses half of the mass in stars bound to the halo), $M_{*}$.


\subsection{Dependence on total halo mass}\label{sec:res:totmass}

\begin{figure}[t!]
        \centering
        \includegraphics[width=\textwidth]{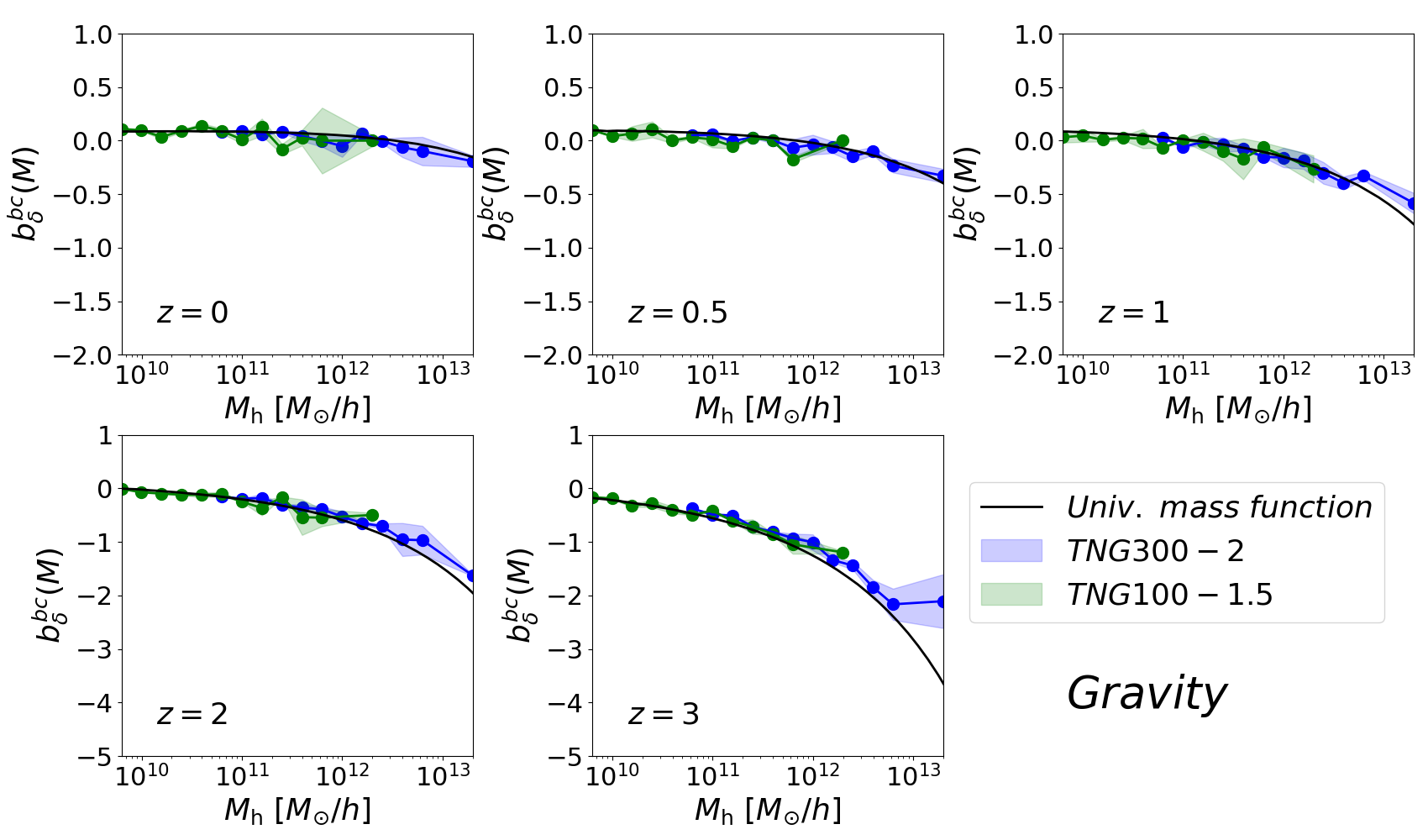}
        \caption{Baryon-CDM density galaxy bias parameter $b_{\delta}^{bc}$ measured as a function of total halo mass in the gravity-only runs, for the resolutions TNG100-1.5 (green) and TNG300-2 (blue) and at different redshifts (different panels), as labeled. The shaded areas bracket the $b_{\delta}^{bc, \rm \high}$ and $b_{\delta}^{bc, \rm \loww}$ values (cf.~Eqs.~(\ref{eq:bcb_measure2}) and \eqref{eq:bcb_measure3}); the dots joined by the solid line indicate their mean (cf.~Eq.~(\ref{eq:bcb_measure})). The black line shows the prediction obtained using the universal mass function formulae (cf.~Eq.~(\ref{eq:tinker_bcb})). Note the different $y$-axis range in the upper and lower panels.}
\label{fig:bcb_totmass_dmo}
\end{figure}

\begin{figure}[t!]
        \centering
        \includegraphics[width=\textwidth]{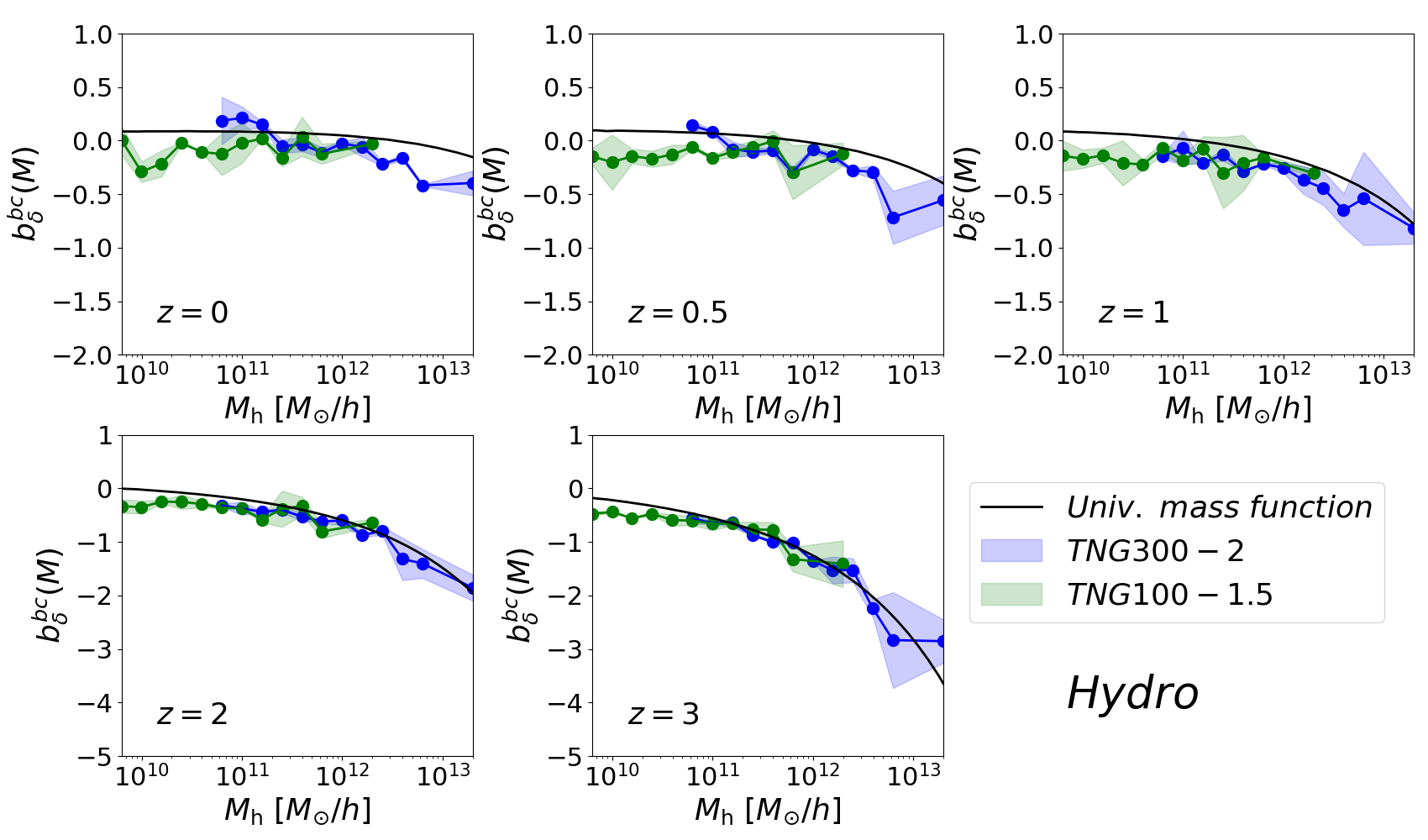}
        \caption{Same as Fig.~\ref{fig:bcb_totmass_dmo}, but for the Hydro runs. The prediction obtained using universal mass function formulae (solid black line) is the same as in Fig.~\ref{fig:bcb_totmass_dmo}.}
\label{fig:bcb_totmass_hydro}
\end{figure}

Figure \ref{fig:bcb_totmass_dmo} shows the baryon-CDM density bias parameter $b_\delta^{bc}$ as a function of total halo mass $M_{\rm h}$ measured from the Gravity runs. The result is shown for the TNG100-1.5 and TNG300-2 resolutions, and for different redshifts, as labeled. Our results show that, overall, $b_\delta^{bc}$ is a decreasing function of halo mass (it becomes more negative with $M_{\rm h}$) and that the amplitude of the effect is stronger at higher redshift. More specifically, $b_\delta^{bc}$ is always negative at $z > 2$ for the mass scales shown, being close to zero for $M_{\rm h} \approx 10^{10}\ M_{\odot}/h$ at both $z=2$ and $z=3$. For halo masses $M_{\rm h} \approx 10^{13}\ M_{\odot}/h$, we find $b_{\delta}^{bc} \approx -1.5\ (z=2)$ and $b_{\delta}^{bc} \approx -3\ (z=3)$. At lower redshifts $z<1$, the mass dependence becomes less pronounced and the amplitude gets overall closer to zero: $b_{\delta}^{bc} \approx -0.3$ and $b_{\delta}^{bc} \approx -0.5$ for $M_{\rm h} \approx 10^{13} M_{\odot}/h$ at redshifts $z=0.5$ and $z=1$, respectively.

The result shown in Fig.~\ref{fig:bcb_totmass_dmo} comes from the gravity-only runs and it therefore captures only the effect due to the modified shape of the linear matter power spectrum (cf.~effect (1) in Sec.~\ref{sec:sims:theory}). This makes it possible to predict it using semi-analytical universal halo abundance formulae such as the Tinker et al.~halo mass function \cite{2008ApJ...688..709T}, in which the number density of halos in a given mass bin $\left[M_{\rm min}, M_{\rm max}\right]$ is given by
\bq\label{eq:tinker_mf}
n(M) &=& \int_{M_{\rm min}}^{{M_{\rm max}}} {\rm d}M' \frac{{\rm d}n(M')}{{\rm d}M'}, \\
\frac{{\rm d}n(M)}{{\rm d}M} &=& f^{\rm T}(\sigma) \frac{\bar{\rho}_{m0}}{M} \frac{{\rm dln}\sigma^{-1}}{{\rm d}M}, \\
f^{\rm T}(\sigma) &=& A\left[\left(\frac{\sigma}{b}\right)^{-a} + 1\right] {\rm exp}\left[-c/\sigma^2\right],
\eq
where $\bar{\rho}_{m0} = 3 \Omega_{m0}H_0^2/(8\pi G)$ is the total physical matter density today and $A = 0.186$, $a = 1.47$, $b=2.57$, $c = 1.19$ are parameters fitted to $\Lambda{\rm CDM}$ gravity-only simulations at $z=0$ for spherical-overdensity halos with mass definition $M_{\rm 200}$ (see Tab.~2 of Ref.~\cite{2008ApJ...688..709T}); the superscript ${}^{\rm T}$ refers to Tinker. In the equations above, $\sigma$ is the variance of the total density field defined as
\bq\label{eq:sigma}
\sigma^2 = \frac{1}{2\pi^2}\int {\rm d}k\,k^2 P_{\delta_m\delta_m}(k,z) \tilde{W}^2(k, R(M)),
\eq
where $\tilde{W}(k, R(M)) = 3\left({\rm sin}(kR) - kR{\rm cos}(kR)\right)/\left(kR\right)^3$ and $R(M) = \left(3M/(4\pi\bar{\rho}_{m0})\right)^{1/3}$. The linear matter power spectrum $P_{\delta_m\delta_m}(k)$ is the only ingredient in the expressions above that depends on the relative abundance of $\Omega_b$ and $\Omega_c$. We can thus predict the baryon-CDM density bias parameter using the universal mass function as 
\bq\label{eq:tinker_bcb}
b_{\delta}^{bc, {\rm univ.}}(z,M) = \frac{1}{\delta_{bc}}\left[\frac{n^{{\rm SepUni}, {\rm univ.}}(z,M)}{n^{{\rm Fiducial}, {\rm univ.}}(z,M)} - 1\right],
\eq
where $n^{\rm Fiducial, univ.}(z,M)$, $n^{\rm SepUni, univ.}(z,M)$ are the universal halo mass function predictions computed with the Tinker fitting function and the linear matter power spectrum of the fiducial and separate universe cosmologies, respectively.  The prediction of Eq.~(\ref{eq:tinker_bcb}) is depicted by the black solid line in Fig.~\ref{fig:bcb_totmass_dmo}, which shows a very good agreement with the simulation results at all of the redshifts and mass scales shown; the larger scatter of the simulation results at higher masses simply reflects the decreased statistical precision due to the smaller number of objects with those masses.  The result is also in line with the physical expectation that a boost in $\Omega_b$ (at fixed $\Omega_m$), i.e., positive $\delta_{bc}$, lowers the amplitude of the linear matter power spectrum on scales $k \gtrsim 0.02\ h/{\rm Mpc}$, which suppresses the formation of the most massive objects. {Recall that the linear power spectrum was obtained assuming constant $\delta_{bc}$ at all times, and thus this result corresponds to super-sound horizon $\delta_{bc}$ modes generated during inflation. For the case of $\delta_{bc}$ modes generated by baryon-photon interactions, $b_{\delta}^{bc}$ is expected to be smaller in absolute value (i.e., less negative) because the suppression in the amplitude of the linear power spectrum for $k \gtrsim 0.02\ h/{\rm Mpc}$ is not as pronounced (cf.~App.~\ref{app:baryon-CDM-theory}).}

Figure \ref{fig:bcb_totmass_hydro} shows the baryon-CDM density bias $b_{\delta}^{bc}$ as a function of total mass $M_{\rm h}$, but now measured from the Hydro runs. The result shown is now due to both the changes in the shape of the linear matter power spectrum (effect (1) in Sec.~\ref{sec:sims:theory}) and the modified hydrodynamics and star formation processes that follow from the different amount of baryons (effect (2) in Sec.~\ref{sec:sims:theory}). In Fig.~\ref{fig:bcb_totmass_hydro}, the universal mass function result (solid black) is the same as in Fig.~\ref{fig:bcb_totmass_dmo}, where it is shown to agree very well with the Gravity results. Thus, comparing the Hydro results with the universal mass function prediction in Fig.~\ref{fig:bcb_totmass_hydro} allows to visualize the impact of effect (2) on the baryon-CDM density bias parameter. Qualitatively, the values of $b_{\delta}^{bc}$ measured from the Hydro runs show a mass and redshift dependence that is very similar to that in the gravity-only results. Quantitatively, the impact of hydrodynamical processes in IllustrisTNG makes the baryon-CDM density bias slightly more negative. This is noticeable at $z = 3$ ($z=2$) for $M_{\rm h} \lesssim 10^{11}\ M_{\odot}/h$ ($M_{\rm h} \lesssim 10^{12}\ M_{\odot}/h$), and effectively all mass scales shown at $z < 1$. For instance, at $z=0.5$ for $M_{\rm h} = 10^{13}\ M_{\odot}/h$, the value of $b_{\delta}^{bc}$ is reduced from $\approx -0.3$ in the Gravity run to $\approx -0.55$ in the Hydro run (the noise in the measurement also increases from the Gravity to the Hydro runs). The suppression in the size of $b_\delta^{bc}$ (i.e., more negative) caused by the hydrodynamical processes suggests that the increased amount of baryons ($\delta_{bc} > 0$) effectively results in amplified feedback effects that suppress overall the number of objects that form at a given total mass $M_{\rm h}$. Finally, we note also that the TNG100-1.5 and TNG300-2 Hydro results are in good agreement, which indicates that our $b_{\delta}^{bc}$ measurements as a function of total halo mass are not strongly affected by limited numerical resolution. 


\subsection{Dependence on galaxy stellar mass}\label{sec:res:stemass}

We now turn our attention to the stellar mass dependence of the baryon-CDM density bias parameter. This measurement has to be made with some care since, at fixed total halo mass, different resolutions return slightly different stellar mass values. This is explored in detail in Ref.~\cite{Pillepich:2017fcc} (see their Appendix A), in which the authors devise a resolution correction strategy to ensure a more trustworthy comparison of quantities that use stellar masses (like the stellar mass function) at different IllustrisTNG resolutions. Specifically, Ref.~\cite{Pillepich:2017fcc} proposes using the ratio of stellar masses found at TNG100-1 and TNG100-2 resolutions as a multiplicative correction factor of the stellar masses found at TNG300-1 resolution, all at fixed total halo mass (see Eq.~(A1) in Ref.~\cite{Pillepich:2017fcc}).\footnote{Important for this to work is the fact that TNG300-1 ($L_{\rm box} = 205\ {\rm Mpc}/h$, $N_p = 2500^3$) and TNG100-2 ($L_{\rm box} = 75\ {\rm Mpc}/h$, $N_p = 910^3$) have approximately the same mass resolution. The higher resolution TNG100-1 simulation has $L_{\rm box} = 75\ {\rm Mpc}/h$, $N_p = 1820^3$.} 

We have applied a similar resolution correction scheme using the original TNG100-2 ($L_{\rm box} = 75\ {\rm Mpc}/h$, $N_p = 910^3$) and TNG100-3 ($L_{\rm box} = 75\ {\rm Mpc}/h$, $N_p = 455^3$) simulations of the \fidu\ cosmology to scale the stellar masses of our TNG300-2 simulations to values representative of TNG100-2 resolutions. Our resolution correction scheme in described in App.~\ref{app:stecorrection}, which shows also that the bias measured from the corrected TNG300-2 catalogues is nearly the same as that measured from the uncorrected ones (cf.~right panel of Fig.~\ref{fig:rtng300}). The reason is that our results are sensitive to relative differences between cosmologies with different $\Omega_b$, $\Omega_c$ values, which are less affected by numerical resolution issues compared to absolute values. In this section, we thus opt to show the results measured from our TNG300-2 simulations without any correction.


\begin{figure}[t!]
        \centering
        \includegraphics[width=\textwidth]{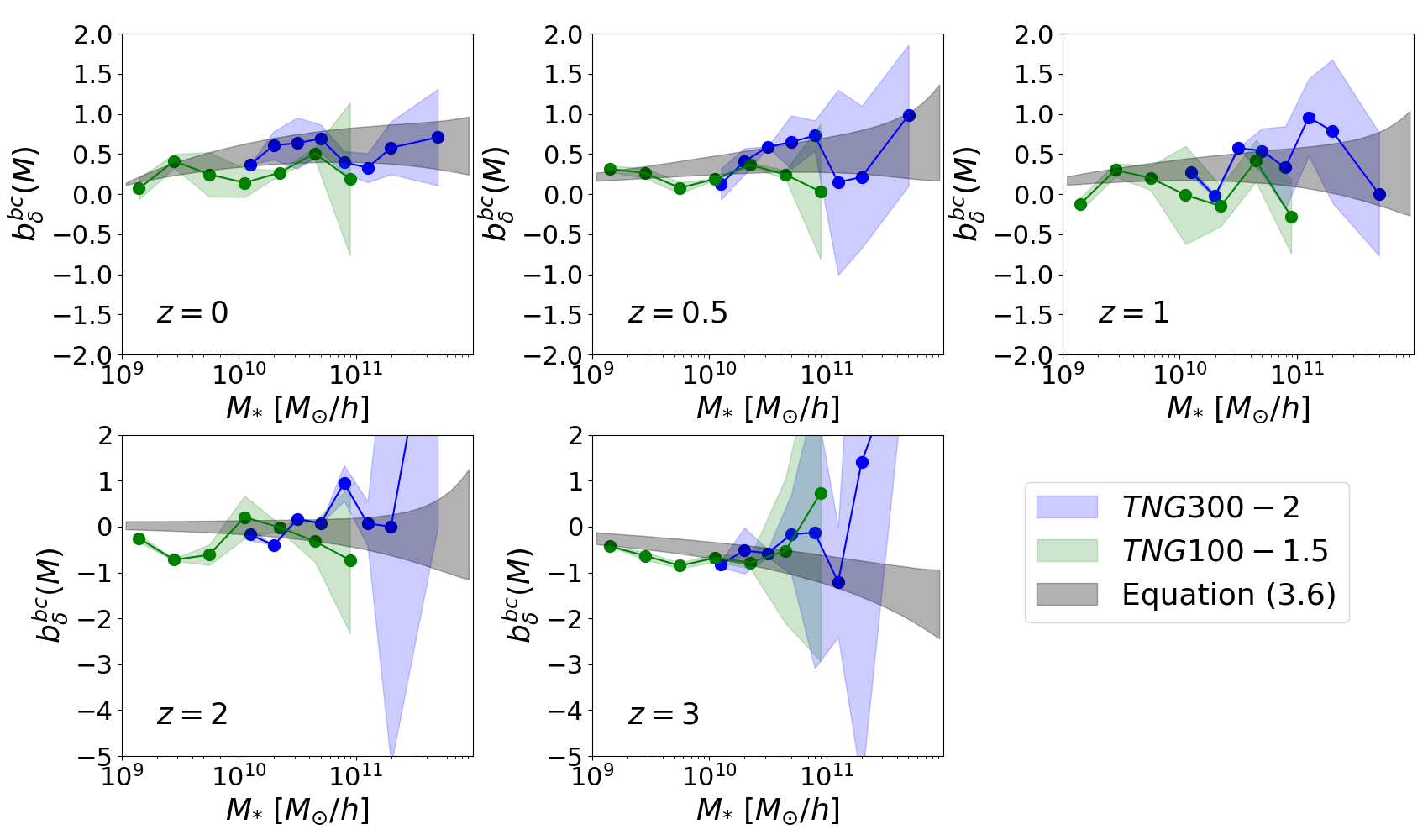}
        \caption{Baryon-CDM density galaxy bias parameter $b_{\delta}^{bc}$ measured as a function of stellar mass $M_{*}$, for the resolutions TNG100-1.5 and TNG300-2 and at different redshifts, as labeled. This is effectively the same as Fig.~\ref{fig:bcb_totmass_hydro}, but as a function of $M_{*}$ instead of $M_{\rm h}$. The grey band shows the result of an approximate model based on the universal mass function and the halo-to-stellar-mass relation of the Fiducial cosmology (cf.~Eq.~(\ref{eq:tinker_bcb_stemass})).}
\label{fig:bcb_stemass_hydro_uncorrected}
\end{figure}

Figure \ref{fig:bcb_stemass_hydro_uncorrected} shows the stellar mass dependence of the baryon-CDM density bias parameter, $b_{\delta}^{bc}(M_{*})$. The figure shows that the two resolutions are in good agreement, although to a slightly lesser extent compared to the same measurement made in terms of total halo mass (cf.~Fig.~\ref{fig:bcb_totmass_hydro}). The difference between $b_{\delta}^{bc, {\rm High}}$ and $b_{\delta}^{bc, {\rm Low}}$ (shaded area) is also larger, but one can nonetheless discern the main differences relative to the measurement made in terms of $M_{\rm h}$. Namely, the shape of $b_{\delta}^{bc}(M_{*})$ is appreciably flatter, with the simulation results suggesting even a slight increase with $M_{*}$ at $z<1$, as opposed to the monotonic decrease with $M_{\rm h}$ displayed in Fig.~\ref{fig:bcb_totmass_hydro}. Further, at redshifts $z < 1$, our results are consistent with $b_{\delta}^{bc}(M_{*})$ being positive for most of the mass scales shown, whereas $b_{\delta}^{bc}(M_{\rm h})$ is always negative in Fig.~\ref{fig:bcb_totmass_hydro}.

\begin{figure}[t!]
        \centering
        \includegraphics[width=\textwidth]{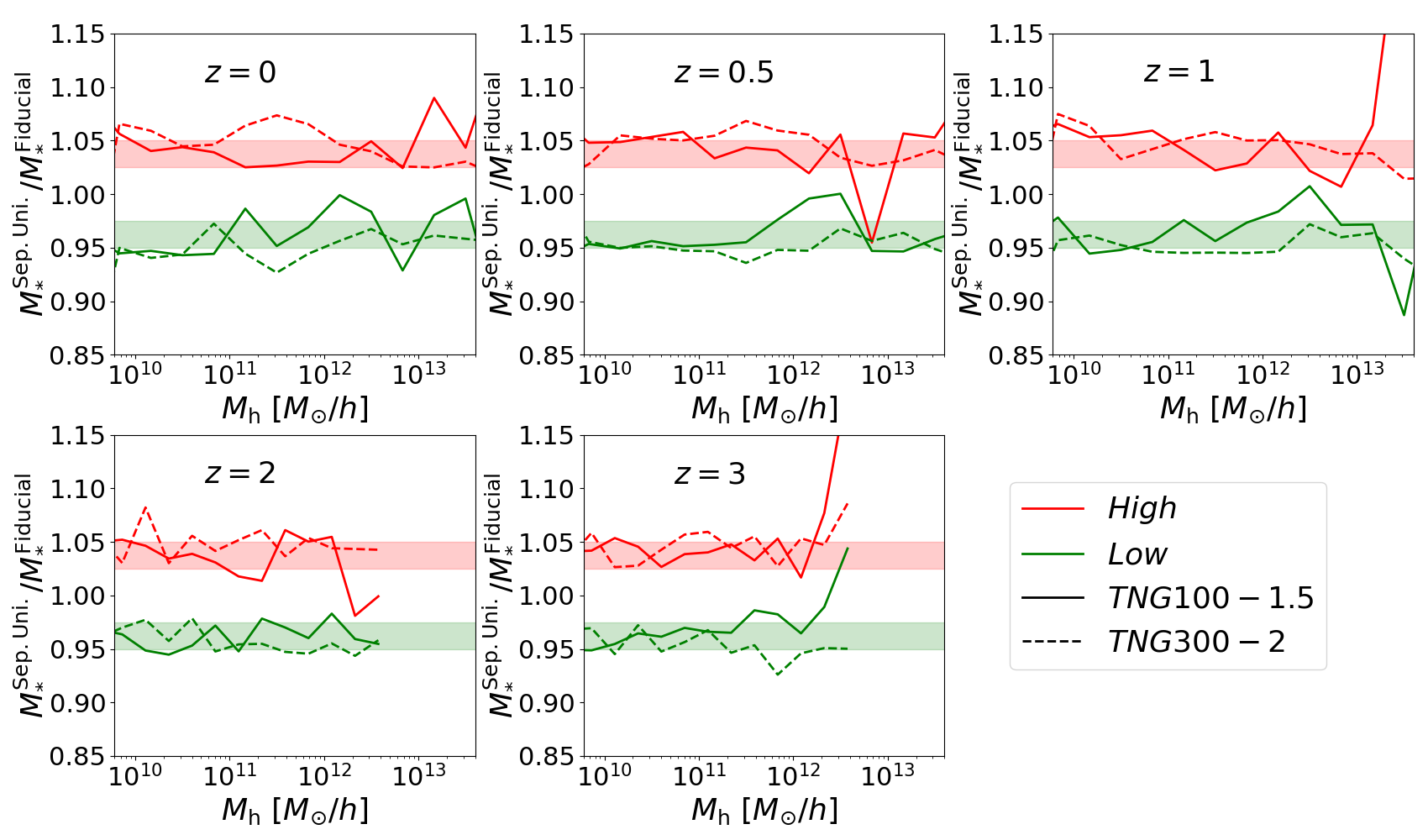}
        \caption{Ratio of the galaxy stellar masses found in the \high\ (red) and \loww\ (green) cosmologies to those found in the \fidu\ cosmology, as a function of total halo mass. The stellar mass values correspond to the median of all values found in bins of total halo mass. The result is shown for the TNG100-1.5 (solid) and TNG300-2 (dashed) resolutions and for different redshifts, as labeled. The horizontal bands bracket changes of $\pm \left[0.5, 1\right] \times \Delta_b$, which correspond roughly to the changes observed in the simulations.}
\label{fig:stemass_ratio_to_fidu}
\end{figure}

As we describe next, the shape of $b_{\delta}^{bc}(M_{*})$ shown in Fig.~\ref{fig:bcb_stemass_hydro_uncorrected} can be explained with the aid of a simple model based on universal mass function formulae and the halo-to-stellar-mass relation of the Fiducial cosmology. Figure \ref{fig:stemass_ratio_to_fidu} shows the impact that changes in $\Delta_b$ have on galaxy stellar masses as a function of halo mass. Specifically, the figure shows the ratio of the median stellar mass in total halo mass bins in the \high\ and \loww\ cosmologies to the \fidu\ one.  As expected, at fixed halo mass, an increase (decrease) in the amount of baryons in the \high\ (${\rm \loww}$) cosmology translates into an increase (decrease) in stellar mass. Further, the magnitude of these changes remains approximately constant across the total halo mass values shown; the two resolutions TNG100-1.5 and TNG300-2 also agree on this result. More quantitatively, the horizontal bands bracket $M_{\rm h}$-independent changes of $50\%$ to $100\%$ of the values of $\Delta_b$ ($\pm \left[0.025, 0.05\right]$), which corresponds roughly to the changes observed in the simulations. This reflects the physical expectation that star formation efficiency is not entirely independent of the cosmic baryon density, and consequently not all of the extra baryons are turned into stars. 

The result of Fig.~\ref{fig:stemass_ratio_to_fidu} indicates that if $M_{\rm h}^{\rm \fidu} \left(M_{*}\right)$ is the median halo-to-stellar-mass relation in the \fidu\ cosmology, then the same relation in the separate universe cosmologies can be approximated as $M_{\rm h}^{\rm Sep.Uni.}(M_*) \approx M_{\rm h}^{\rm \fidu} \left(\left[1 - \epsilon\Delta_b\right]M_{*}\right)$ where $\epsilon$ is a number between $0.5$ and $1$. One can then plug these halo-to-stellar-mass relations into the universal mass function formulae and generalize Eq.~(\ref{eq:tinker_bcb}) as
\bq\label{eq:tinker_bcb_stemass}
b_{\delta}^{bc, {\rm univ.}}(z,M_{*}) = \frac{1}{\delta_{bc}}\left[\frac{n^{{\rm SepUni}, {\rm univ.}}\left(z,M_{\rm h}^{\rm \fidu} \left(\left[1 - \epsilon\Delta_b\right]M_{*}\right)\right)}{n^{{\rm \fidu}, {\rm univ.}}\left(z,M_{\rm h}^{\rm \fidu} \left(M_{*}\right)\right)}- 1\right].
\eq
This equation's prediction is shown by the shaded grey band in Fig.~\ref{fig:bcb_stemass_hydro_uncorrected}, which brackets the result from using $\epsilon = 0.5$ and $\epsilon = 1$. We evaluate $M_{\rm h}^{\rm \fidu} \left(M_{*}\right)$ using a polynomial fit to the median relations found in the TNG100-1.5 simulation of the Fiducial cosmology. Figure \ref{fig:bcb_stemass_hydro_uncorrected} shows that the simple model of Eq.~(\ref{eq:tinker_bcb_stemass}) succeeds at explaining the overall amplitude of the $b_{\delta}^{bc} (z,M_{*})$ measured in the simulations, as well as its dependence on $z$ and $M_{*}$. The level of agreement is not perfect, but this is as expected for at least three reasons: (i) the universal mass function with the Tinker fitting formula is already not a perfect description of the halo mass function in hydrodynamical simulations (cf.~Fig.~\ref{fig:bcb_totmass_dmo} vs.~Fig.~\ref{fig:bcb_totmass_hydro}); (ii) the relation $M_{\rm h}^{\rm \fidu} \left(M_{*}\right)$ we use is fitted to the median total halo mass found in a given stellar mass bin, which fails to capture the shape of the distribution within the bin; (iii) we have assumed that $\epsilon$ is constant with mass, which is an approximation valid only to the degree \mbox{shown in Fig.~\ref{fig:stemass_ratio_to_fidu}.} 

\begin{figure}[t!]
        \centering
        \includegraphics[width=\textwidth]{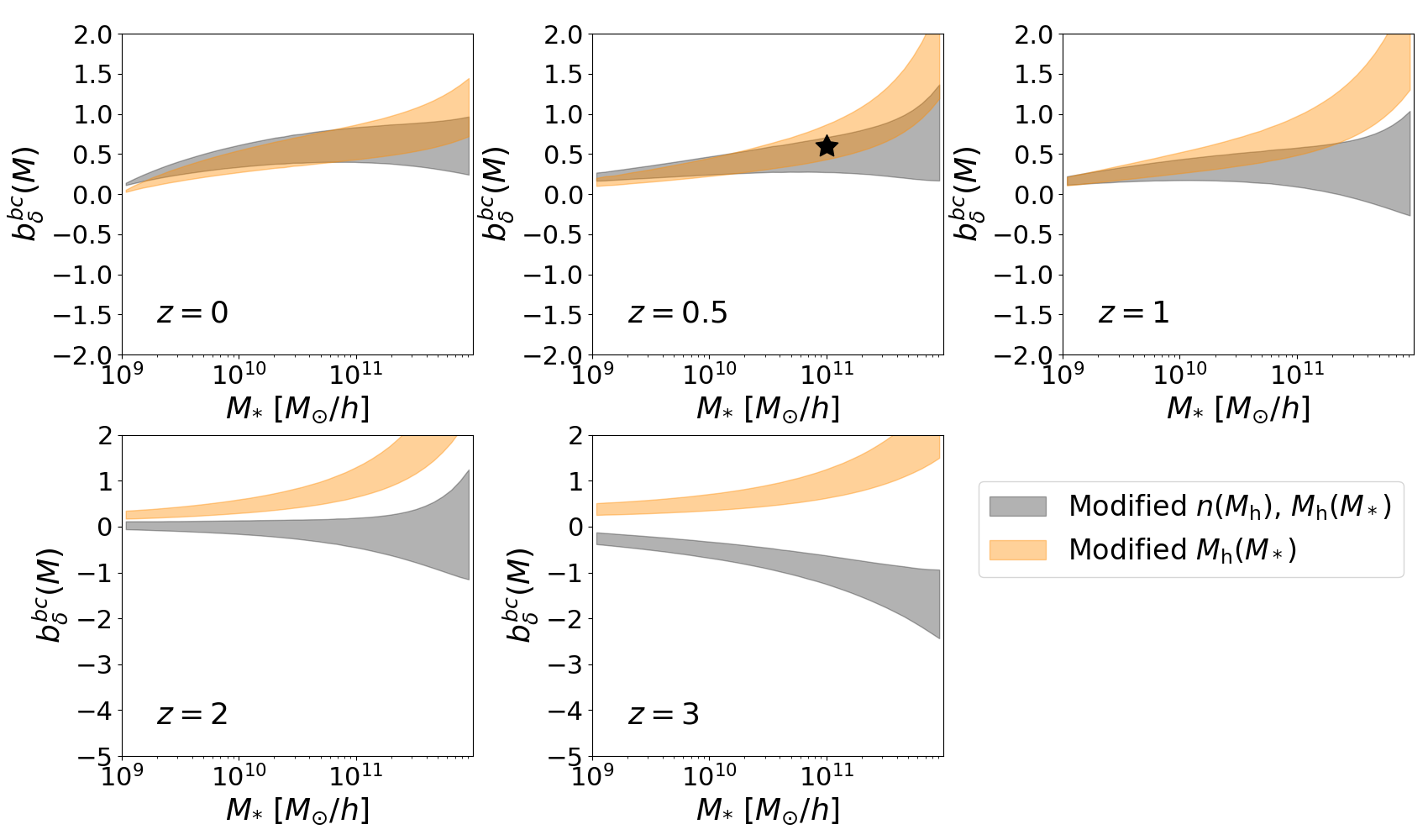}
        \caption{{Comparison of the relative contribution to $b_{\delta}^{bc}$ from the modifications to the halo abundances and the stellar-to-halo-mass relations. The grey band brackets the result of Eq.~(\ref{eq:tinker_bcb_stemass}) for $\epsilon\in\left[0.5, 1\right]$ and it corresponds to the combined effect of modified halo mass functions and stellar-to-halo-mass relations (i.e., including effects (1) and (2); it is the same as that shown in Fig.~\ref{fig:bcb_stemass_hydro_uncorrected}). The orange band shows the same quantity, but using $n^{{\rm \fidu}, {\rm univ.}}$ instead of $n^{{\rm SepUni}, {\rm univ.}}$ in the numerator of Eq.~(\ref{eq:tinker_bcb_stemass}); it thus captures only effect (2), the effect of a modified $\Omega_b/\Omega_c$ on the stellar-to-halo-mass relation. The star in the $z=0.5$ panel indicates the value adopted in Sec.~\ref{sec:Pgk} to estimate the impact of $\delta_{bc}$ on the power spectrum of BOSS galaxies.}}
\label{fig:bcb_stemass_hydro_modelsonly}
\end{figure}

{To gain more insight on how much of the shape of $b_{\delta}^{bc}(M_*)$ is due to the changes in the halo mass function and the halo-to-stellar-mass relation, we compare in Fig.~\ref{fig:bcb_stemass_hydro_modelsonly} the result of Eq.~(\ref{eq:tinker_bcb_stemass}) (grey band) with that obtained by considering only the changes to the $M_{\rm h}(M_*)$ relation, i.e., by replacing $n^{{\rm SepUni}, {\rm univ.}}$ with $n^{{\rm \fidu}, {\rm univ.}}$ in the numerator of Eq.~(\ref{eq:tinker_bcb_stemass}) (orange band). Noting that the halo abundances are predominantly controlled by the shape of the power spectrum (effect (1) in Sec.~\ref{sec:sims:theory}; cf.~Figs.~\ref{fig:bcb_totmass_dmo} vs.~\ref{fig:bcb_totmass_hydro}), the orange band thus captures effect (2) in Sec.~\ref{sec:sims:theory} in isolation. The figure shows that the lower the redshift, the greater the importance of effect (2). This result could have been anticipated from Figs.~\ref{fig:bcb_totmass_dmo} and \ref{fig:stemass_ratio_to_fidu}, which show that the impact on halo abundances decreases with redshift, but the changes to the stellar-to-halo-mass relations remain effectively constant. Specifically, at $z=0, 0.5$ and $1$, effect (1) has barely any impact for $M_* \lesssim 10^{12}, 10^{11}$ and $10^{10}\ M_{\odot}/h$, respectively. On the other hand, for $z>2$, both effects (1) and (2) contribute sizeably, and $b_{\delta}^{bc}$ is given by the result of their competition.}

{Figure \ref{fig:bcb_stemass_hydro_modelsonly} allows us to also comment on the expected result for $\delta_{bc}$ modes generated by photon-baryon interactions. In the previous section, we noted that the impact on halo abundances registered in our simulations of constant $\delta_{bc}$ modes can be regarded as upper bounds to the impact of $\delta_{bc}$ modes generated by baryon-photon interactions (because the impact on the initial power spectrum is not expected to be as large). Further, the stellar-to-halo-mass relation is sensitive mostly to the relative amount of baryons and CDM at the onset of nonlinear structure formation; in other words, if $\delta_{bc}$ is already a constant at the starting time of the simulation, then the result of Fig.~\ref{fig:stemass_ratio_to_fidu} is independent of the exact past evolution of $\delta_{bc}$. Putting these arguments together, we expect the $b_{\delta}^{bc}$ values for $\delta_{bc}$ generated by photon-baryon interactions to be bracketed by the orange and grey bands. Further, noting that at low redshift the two practically overlap, we conclude that for these redshifts the stellar mass dependence of $b_{\delta}^{bc}$ is fairly independent of the exact origin of the $\delta_{bc}$ modes (inflation vs. baryon-photon interactions).}


\section{Impact on the galaxy power spectrum}\label{sec:Pgk}

We can use the values of the baryon-CDM density galaxy bias parameter $b_{\delta}^{bc}$ measured from the simulations to estimate the corresponding impact on the galaxy power spectrum. We work to linear order, do not include redshift-space distortions and consider the following galaxy bias expansion (we keep the mass dependence of the bias parameters implicit to ease the notation): 
\bq\label{eq:bias_exp_impact}
\delta_g(\vx, z) = b_1(z)\delta_m(\vx, z) + b_\delta^{bc}(z) \delta_{bc}(\vx).
\eq
We focus only on the baryon-CDM density contribution, i.e.~we disregard the contribution from the relative baryon-CDM velocity divergence term $\theta_{bc}(\vx, z)$ (whose bias parameter we cannot measure with our separate universe simulations). After Fourier-transforming, the galaxy power spectrum $P_{gg}(k,z)$ can be written as
\bq\label{eq:pgg_impact}
P_{gg}(k,z) &=& b_1(z)^2 P_{\delta_m\delta_m}(k, z) + 2b_1(z)b_{\delta}^{bc}(z)P_{\delta_m\delta_{bc}}(k, z) + b_{\delta}^{bc}(z)^2P_{\delta_{bc}\delta_{bc}}(k, z),
\eq
where $P_{\delta_m\delta_{bc}}$ is the cross power spectrum of $\delta_m$ and $\delta_{bc}$ and $P_{\delta_{bc}\delta_{bc}}$ is the auto power spectrum of $\delta_{bc}$.  {In this section, we assume that the $\delta_{bc}$ modes are generated by photon-baryon interactions before decoupling; the calculation of the corresponding spectra is described in App.~\ref{app:baryon-CDM-theory}. In order to evaluate $P_{gg}(k,z)$ using Eq.~(\ref{eq:pgg_impact}), we need to specify the values of $b_1$ and $b_\delta^{bc}$. We consider galaxy samples like that of BOSS DR12 with typical stellar masses of order $M_* \approx 10^{11}\ M_{\odot}/h$ \cite{2016MNRAS.460.1457S} at median effective redshifts $z_{\rm eff} \approx 0.5$. The value of $b_\delta^{bc} = 0.6$ can be read off from the $z = 0.5$ panel of Fig.~\ref{fig:bcb_totmass_hydro} (black star). For $b_1$ we use the halo bias semi-analytical formulae of Ref.~\cite{2010ApJ...724..878T} assuming a typical host halo mass for BOSS galaxies with $M_h \approx 10^{13}\ M_{\odot}/h$ (in accordance with halo abundance matching analyses \cite{2016MNRAS.460.1457S}). This yields $b_1 = 1.5$, which is also in line with the constraints on $b_1\sigma_8 \approx 1.3$ obtained by Ref.~\cite{2017MNRAS.466.2242B} for the same sample.}

\begin{figure}[t!]
        \centering
        \includegraphics[width=\textwidth]{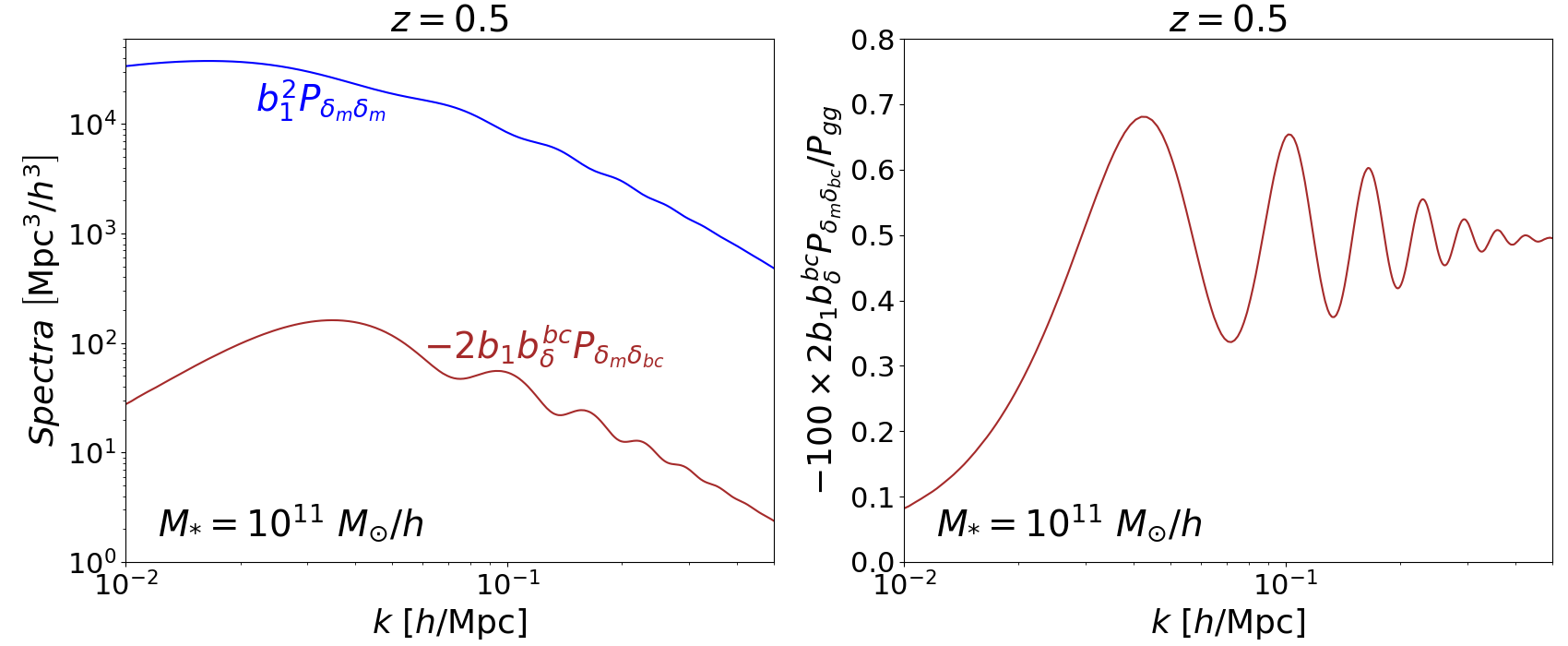}
        \caption{{Contribution of the baryon-CDM density bias $b_{\delta}^{bc}$ to the galaxy power spectrum at leading order (brown lines), for bias values that are representative of BOSS DR12-like galaxies at $z \approx 0.5$: $b_{\delta}^{bc} = 0.6$ and $b_1 = 1.5$ ($M_* \approx 10^{11}\ M_{\odot}/h$ and $M_h \approx 10^{13}\ M_{\odot}/h$). $P_{\delta_m\delta_{bc}}$ is the cross spectrum between $\delta_m$ and $\delta_{bc}$ (assumed generated by photon-baryon interactions), which is negative. The left panel shows the absolute values and the right panel shows the relative contribution (in percent). The linear LIMD bias contribution ($\propto b_1^2$; blue) is also shown.}}
\label{fig:pgk}
\end{figure}

{The left panel of Fig.~\ref{fig:pgk} shows the contribution from the $P_{\delta_m\delta_m}(k, z)$ and $P_{\delta_m\delta_{bc}}(k,z)$ terms in Eq.~(\ref{eq:pgg_impact}); the right panel shows the relative contribution of the $P_{\delta_m\delta_{bc}}(k,z)$ term to the total $P_{gg}(k,z)$. Overall, the figure makes apparent that the $b_1^2P_{\delta_m\delta_m}$ term is the dominant contribution for BOSS DR12-like galaxies, with the $P_{\delta_m\delta_{bc}}(k,z)$ accounting for a subpercent fraction that oscillates between $0.3\%$ and $0.7\%$ for the scales shown. The contribution from the term $\propto P_{\delta_{bc}\delta_{bc}}(k,z)$ in Eq.~(\ref{eq:pgg_impact}) is not shown, but we have explicitly confirmed that it is smaller in absolute value than the $P_{\delta_m\delta_{bc}}(k,z)$ term by over two orders of magnitude.}

{The relatively small impact of the baryon-CDM density perturbations on the galaxy power spectrum can be understood by inspecting the ratio of the $P_{\delta_m\delta_{bc}}$ to the $P_{\delta_m\delta_m}$ contributions in Eq.~(\ref{eq:pgg_impact}), which, in the case of adiabatic primordial perturbations considered in this section, is given by
\bq\label{eq:ratio_impact}
\frac{2b_1(z)b_{\delta}^{bc}(z)P_{\delta_m\delta_{bc}}(k, z)}{b_1(z)^2 P_{\delta_m\delta_m}(k, z)} \approx \frac{2b_{\delta}^{bc}(z)}{b_1(z)} \frac{T_b(k,z) - T_c(k,z)}{T_m(k,z)},
\eq
where $T_b$, $T_c$ and $T_m$ are the transfer functions of baryons, CDM and total matter, respectively (defined as $\delta_s(k, z) = (2k^2/(5\Omega_mH_0^2))\R(k)T_s(k, z)$, $s \in \{b, c, m\}$, where $\R(k)$ is the adiabatic primordial scalar perturbation generated by inflation); the approximation in Eq.~(\ref{eq:ratio_impact}) is valid at late times, as explained in App.~\ref{app:baryon-CDM-theory}. The main suppression is due to the small absolute value of $(T_b - T_c)/T_m$, which Ref.~\cite{2016PhRvD..94f3508S} had used to predict already $\approx 1\%$-level contributions using $b_{\delta}^{bc} = 1$ (see also Ref.~\cite{barkana/loeb:11} for a similar earlier prediction and Ref.~\cite{2019JCAP...06..006C} for the impact of larger bias values). An additional small suppression comes from the ratio $2b_{\delta}^{bc}/b_1 \approx 0.8$. Also, note that the bias parameters are positive (for stellar mass selection at $z=0.5$), but $T_b - T_c$ is negative, and hence, the baryon-CDM density perturbations contribute negatively to the galaxy power spectrum. Further, baryon-CDM perturbations also contribute to the cross spectrum of galaxies with total matter (which is relevant for galaxy-galaxy lensing measurements), but their relative contribution is suppressed by a similar amount following the same arguments.}

{The oscillatory behavior of the relative contribution of $P_{\delta_m\delta_{bc}}(k,z)$ is indicative of an offset of the phases of the BAO oscillations. In configuration space, i.e.~at the level of the galaxy two-point correlation function, this can potentially lead to shifts in the position of the BAO peak, which is one of the main geometrical probes of the expansion history of the Universe. References~\cite{barkana/loeb:11, 2016PhRvD..94f3508S} estimated that the BAO peak position is likely shifted by less than $1\%$ by the baryon-CDM density term. Our results sharpen this conclusion, since we now have a likely range for the bias parameter $b_{\delta}^{bc}$. Figure 5 of Ref.~\cite{2017MNRAS.470.2723B} shows the expected biases on the distance and Hubble-rate parameters $\alpha_{\perp}$ and $\alpha_{\parallel}$, and growth rate of structure $f\sigma_8$ that would arise from ignoring $\delta_{bc}$ in observational analyses of the BOSS DR12 galaxy sample.\footnote{These parameters are defined as
\bq
\alpha_\perp = \frac{D_A(z)r^*_d}{D^*_A(z)r_d} \ \ \ ; \ \ \ \alpha_\parallel = \frac{H^*(z)r^*_d}{H(z)r_d}  \ \ \ ; \ \ \  f\sigma_8(z) = \frac{{\rm dln}D(z)}{{\rm dln}a}\sigma_8(z),
\eq
where $D_A^*(z)$ is the angular diamater distance to redshift $z$, $r^*_d$ is the sound horizon at photon-baryon decoupling and $H^*(z)$ is the Hubble rate, all evaluated in a reference cosmology assumed in the analysis; the same quantities without the superscript $^*$ denote the true/unknown values that are to be inferred from the data. Further, $D(z)$ is the linear growth factor of total matter perturbations.} At $95\%$ confidence level, the analysis of Ref.~\cite{2017MNRAS.470.2723B} constrains $b_{\delta}^{bc} = -1.0 \pm 6.2$, which keeps the biases on $\alpha_{\perp}$, $\alpha_{\parallel}$ and $f\sigma_8$ below the $0.5\%$, $0.5\%$ and $2\%$ levels, respectively (cf.~left middle panel of Fig.~5 of Ref.~\cite{2017MNRAS.470.2723B}). Our simulation results show that $b_{\delta}^{bc} \approx 0.6$ for galaxy stellar masses and redshifts representative of BOSS DR12. Taking the estimates of Ref.~\cite{2017MNRAS.470.2723B} as a guide, the expected biases caused by our $\delta_{bc}$ values are reduced to $\approx 0.1\%$ for all $\alpha_{\perp}$, $\alpha_{\parallel}$ and $f\sigma_8$. Furthermore, according to the forecasts of Ref.~\cite{2019JCAP...06..006C}, the bias values measured from our simulations ($-1\lesssim b_{\delta}^{bc} \lesssim 1$; cf.~Fig.~\ref{fig:bcb_stemass_hydro_modelsonly}) should remain below $1\sigma$ detection thresholds in surveys like DESI \cite{2013arXiv1308.0847L}.}

{Before summarizing, we note that there is an interesting application of our results in the context of constraints on primordial compensated isocurvature perturbations (CIP) generated during inflation (see e.g.~Refs.~\cite{2009PhRvD..80f3535G, 2010ApJ...716..907H, grin/dore/kamionkowski, 2014PhRvD..89b3006G, 2017PhRvD..96h3508S, 2016PhRvD..93d3008M, 2019MNRAS.485.1248S, 2019arXiv190400024H, 2019arXiv190808953H} for a number of existing observational constraints and forecasts). These} contribute to the galaxy power spectrum as $\propto b_1b_{\delta}^{bc}A_{\rm CIP} P_{\delta_m\R}(k, z)$, where $A_{\rm CIP} = \Delta_b / \R$ is a parameter that determines the size of isocurvature perturbations $\Delta_b$ relative to the adiabatic curvature perturbations $\R$. This contribution has the same scale dependence as the one coming from primordial non-Gaussianity of the local type, whose amplitude is parametrized by the $f_{\rm NL}$ parameter \cite{2001PhRvD..63f3002K}. {Noting that our measurements give $b_{\delta}^{bc}$ values that are $\mathcal{O}(0.1 - 1)$ (the sign and exact value can depend on redshift and mass selection)}, it follows that galaxy surveys like SPHEREx \cite{2014arXiv1412.4872D} that aim to place $\O(1)$ constraints on $f_{\rm NL}$ should be able to place similarly tight constraints on $A_{\rm CIP}$, since both $f_{\rm NL}$ and $A_{\rm CIP}$ impact the galaxy power spectrum in the same way. More recently, Ref~\cite{2019arXiv190808953H} presented also forecast constraints on CIP from contributions $\propto b_{\delta}^{bc}A_{\rm CIP}$ in the tomographic kinetic Sunyaev-Zel'dovich signal.


\section{Summary and conclusions}\label{sec:conc}

In this paper, we have used hydrodynamical cosmological simulations to study the impact of baryon-CDM perturbations on galaxy formation. {These are compensated isocurvature perturbations between baryons and CDM that can be generated during inflation if multiple fields are present. However, even for adiabatic initial conditions after inflation, these perturbations will also be produced} during the epoch prior to baryon-photon decoupling, when baryons and CDM did not comove because of the tight coupling of the baryons to the photons (and lack thereof for CDM). Once the baryons decoupled from the photons, they were able to collapse gravitationally, but there were regions within which the amount of baryons relative to CDM differed from that at cosmic mean and baryons were moving with a different velocity relative to the CDM component. These {\it baryon-CDM density and velocity perturbations} can naturally have an impact on galaxy formation, but are not customarily included in hydrodynamical simulations, as well as in theoretical models of galaxy clustering statistics. Importantly, strong BAO features are imprinted on the statistics of the baryon-CDM perturbations, so assessing their contribution to galaxy clustering is important to ensure robust cosmology inference from the BAO feature.

We have focused on long-wavelength compensated baryon-CDM density perturbations $\delta_{bc}$ characterized by $\delta_c = -f_b\delta_b$ and $\delta_m = 0$. Under the separate universe ansatz, galaxy formation taking place sufficiently inside these baryon-CDM density perturbations is effectively equivalent to structure formation taking place in a cosmology with modified baryon ($\Omega_b$) and CDM ($\Omega_c$) density parameters and fixed total matter ($\Omega_m$; cf.~Tab.~\ref{table:params}). There are two ways in which this change in cosmology impacts galaxy formation. The first is that an increased (decreased) value of $\Omega_b/\Omega_c$ lowers (enhances) the linear matter power spectrum on scales $k \gtrsim 0.02\ h/{\rm Mpc}$. {In this paper, we have assumed that the value of $\Omega_b/\Omega_c$ is constant at all times after inflation, which maximizes this suppression on the matter power spectrum. For baryon-CDM perturbations generated by photon-baryon interactions, the long-wavelength perturbations in $\Omega_b/\Omega_c$ are time-dependent before decoupling and are expected to lead to a weaker modulation of the small-scale modes.} The second effect is that modified relative amounts of baryons and CDM also result in different fractions of the total matter that can experience hydrodynamical forces and form stars.

Baryon-CDM density perturbations enter the galaxy bias expansion at leading order (cf.~Eq.~(\ref{eq:bias_exp1})) and our main objective was to measure the corresponding bias parameter $b_{\delta}^{bc}$. We have done so by measuring the response of galaxy number counts to changes in the amplitude of $\delta_{bc}$ (cf.~Eqs.~(\ref{eq:bias_as_response}), \eqref{eq:bcb_measure}, \eqref{eq:bcb_measure2} and \eqref{eq:bcb_measure3}). We have performed our simulations using the {\sc AREPO} code with the IllustrisTNG galaxy formation model at TNG100-1.5 ($L_{\rm box} = 75\ {\rm Mpc}/h$, $N_p = 1250^3$) and TNG300-2 ($L_{\rm box} = 205\ {\rm Mpc}/h$, $N_p = 1250^3$) resolutions. For each resolution, we have run simulations without (referred to as ``Gravity'') and with (``Hydro'') hydrodynamical processes taken into account. We have studied the redshift and mass dependence of $b_{\delta}^{bc}$ for two mass definitions: (i) the total mass due to all mass elements that belong to the host halo of the galaxy, $M_{\rm h}$; (ii) the mass in stars within twice the stellar half-mass radius, $M_{*}$. A summary of our main results is as follows:
\begin{itemize}

\item {The bias parameter $b_{\delta}^{bc}$ becomes more negative with increasing host halo mass  $M_{\rm h}$, and it does so more pronouncedly at higher redshift (cf.~Fig.~\ref{fig:bcb_totmass_hydro}). Quantitatively, we find for $M_{\rm h} = 10^{13}\ M_{\odot}/h$ that $b_{\delta}^{bc} \approx -0.5, -0.7, -1.5$ at $z = 0.5, 1, 2$, respectively. At $z=1$, we have $b_{\delta}^{bc} \approx -0.1, -0.3, -0.7$ for $M_{\rm h} = 10^{11}, 10^{12}, 10^{13}\ M_{\odot}/h$, respectively.}

\item The mass dependence of $b_{\delta}^{bc}$ becomes markedly different when galaxies are selected according to stellar mass $M_{*}$, instead of $M_{\rm h}$ (cf.~Fig.~\ref{fig:bcb_stemass_hydro_uncorrected}). In particular, our results suggest a slight increase of $b_{\delta}^{bc}$ with $M_{*}$ and the redshift dependence is also less pronounced. {At redshifts $z < 1$, $b_{\delta}^{bc}$ is positive for $M_{*} \gtrsim 2\times10^{10}\ M_{\odot}/h$; for $M_* \approx 10^{11}\ M_{\odot}/h$ at $z=0.5$, which is representative of BOSS galaxies, we find $b_{\delta}^{bc} = 0.6$.}

\item  {The values of $b_{\delta}^{bc}$ for both halo-mass- and stellar-mass-selected tracers can be explained with a simple empirical model that takes into account (1) the effect of the modified linear power spectrum on the halo mass function via a universal mass function (Figs.~\ref{fig:bcb_totmass_dmo} and \ref{fig:bcb_totmass_hydro}); and (2) the modification of the stellar mass-halo mass relation  $M_*(M_{\rm h})$ due to the changed baryon fraction (Fig.~\ref{fig:stemass_ratio_to_fidu}). For halo-mass-selected tracers, only the first effect is relevant in this simple model, while the bias $b_{\delta}^{bc}$ of stellar-mass-selected tracers is determined by the combination of both effects, with effect (2) being dominant at low redshifts (cf.~Fig.~\ref{fig:bcb_stemass_hydro_modelsonly}).}

\item The contribution $2b_1b_{\delta}^{bc}P_{\delta_m\delta_{bc}}(k)$ from baryon-CDM density perturbations {generated by baryon-photon interactions} to the galaxy power spectrum is expected to lie below the $1\%$ level for galaxy samples like BOSS DR12 (cf.~Fig.~\ref{fig:pgk}). Even taking into account the stronger BAO feature in $\delta_{bc}$, the impact of $b_{\delta}^{bc}$ is expected to be of order $0.1\%$ on the inferred distance and Hubble-rate parameters $\alpha_{\perp}$, $\alpha_{\parallel}$, and growth rate of structure $f\sigma_8$ (cf.~Sec.~\ref{sec:Pgk}), at least for an analysis pipeline similar to that of Ref.~\cite{2017MNRAS.470.2723B}.

\end{itemize}

The impact of baryon-CDM perturbations is expected to be more pronounced at earlier times because the size of the constant $\delta_{bc}$ mode relative to the growing total matter density $\delta_m$ increases toward higher redshifts. This provides motivation to extend our baryon-CDM density perturbation study to higher redshifts (we limited ourselves to $z<3$ here), which would allow to study the impact of baryon-CDM density perturbations on the formation of the first halos, stars and galaxies. Further, at higher redshift, it would be interesting to go beyond the study of the impact of baryon-CDM perturbations on the total number of galaxies and measure the response of observables associated with the gas distribution. An example is the study of the response of the distribution of neutral Hydrogen, which could provide insights on the epoch of reionization ($z \approx 10 - 7$) and the signals of $21$-cm emission and Lyman-$\alpha$ absorption spectra.


It would also be interesting to extend our implementation of baryon-CDM perturbations in IllustrisTNG to include not only $\delta_{bc}$, but also the relative velocity divergence $\theta_{bc}$ and uniform relative velocity $\vv_{bc}$ terms. Past simulation work focusing on $z \gtrsim 10$ \cite{popa/etal, 2011ApJ...736..147G, 2011MNRAS.412L..40M, 2011ApJ...730L...1S, 2012ApJ...747..128N, Visbal/etal:12, 2012ApJ...760....4O, 2013ApJ...763...27N, Fialkov:2014rba, 2013ApJ...771...81R, 2019ApJ...878L..23C} has shown that the presence of a uniform relative velocity suppresses the formation of low-mass haloes, as well as the rate at which gas can cool and form stars at early times. These simulations require however appreciably higher resolution than the ones currently performed with the IllustrisTNG model. The imprints that these early-time effects can leave on later-time structure formation (e.g., via modified reionization histories) are therefore currently uncertain, which motivates further investigations; we defer these to future work. 

Finally, we note that our baryon-CDM bias measurements are specific to the IllustrisTNG galaxy formation model, and that it is plausible to expect the result to vary for varying implementations of baryon physics in cosmological simulations of galaxy formation. The bias parameters describe formally the environmental dependence of galaxy formation, and hence, a comparison of galaxy bias predictions from different baryon physics models can be used to take interesting steps in studies of galaxy formation.

\begin{acknowledgments}

We thank Raul Angulo, {Simeon Bird, Jonathan Blazek, Emanuele Castorina}, Bruno Henriques, Titouan Lazeyras, Kaloian Lozanov, R\"udiger Pakmor, Annalisa Pillepich, Volker Springel and Simon White for useful comments and discussions. The simulations used in this work were run on the Cobra supercomputer at the Max Planck Computing and Data Facility (MPCDF) in Garching near Munich. AB, GC and FS acknowledge support from the Starting Grant (ERC-2015-STG 678652) ``GrInflaGal'' of the European Research Council.
\end{acknowledgments}

\appendix 

\section{Evolution of baryon-CDM perturbations}\label{app:baryon-CDM-theory}

In this appendix, we present the equations that describe the evolution of baryon-CDM perturbations. We evaluate first the general solution of the baryon-CDM density and velocity divergence perturbations, and then we describe the calculation of the power spectra {for those generated by baryon-photon interactions}.

\subsection{The constant and decaying modes}\label{app:baryon-CDM-theory_1}

After photon-baryon decoupling, the linear density contrast $\delta$ and velocity divergence $\theta = {\bm\nabla} \cdot \vv$ of baryons and CDM obey the following equations
\bq
\label{eq:delta_s} \dot{\delta}_s(\vx, z) &=& -\frac{\theta_s(\vx, z)}{a} \ \ \ \ \ \ \ \ \ \ \ \ \ \ \ \ ,\ \ \ \ s \in \{b, c\} \\
\label{eq:theta_s} \dot{\theta}_s(\vx, z) + H(z)\theta_s(\vx, z) &=& -4\pi G a \bar{\rho}_m(z)\delta_m(\vx, z)  \ , \ \ \ \ s \in \{b, c\},
\eq
where the subscripts $_b$, $_c$ and $_m$ denote baryons, CDM and total matter respectively, $H(z)$ is the Hubble rate, $\bar{\rho}_m(z)$ is the background density of total matter and an overdot denotes a derivative w.r.t.~physical time $t$; we use $t$, the scale factor $a$ and the redshift $z$ interchangeably as time variables. With these equations, we can write the following three equations (not independent) for the relative density, $\delta_r = \delta_b - \delta_c$, and relative velocity divergence $\theta_r = \theta_b - \theta_c$ between baryons and CDM:
\bq
\label{eq:delta_r} \ddot{\delta}_r(\vx, z) + 2H(z)\dot{\delta}_r(\vx, z) &=& 0, \\
\label{eq:theta_r} \dot{\theta}_r(\vx, z) + H(z)\theta_r(\vx, z) &=& 0, \\
\label{eq:other_r} \dot{\delta}_r(\vx, z) &=& - \frac{\theta_r(\vx, z)}{a}.
\eq
From Eq.~(\ref{eq:delta_r}), we know that the solution for $\delta_r$ is of the form (we follow closely the notation of Ref.~\cite{2016PhRvD..94f3508S})
\bq\label{eq:delta_r_sol}
\delta_r(\vx, z) = \delta_{bc}(\vx) + R_-(\vx) D_r(z),
\eq
where $\delta_{bc}(\vx)$ and $R_-(\vx)$ are time-independent and $D_r(z)$ is a decaying function of time that satisfies $\ddot{D}_r + 2H(z)\dot{D}_r = 0$ and admits the following integral solution (the $H_0$ factor merely makes $D_r(z)$ dimensionless)
\bq\label{eq:Dr}
D_r(z) = H_0 \int_{t(z)}^{\infty} \frac{dt'}{a(t')^2}.
\eq
Strictly speaking, $\delta_{bc}(\vx)$ and $R_-(\vx)$ should be evaluated at the Lagrangian position associated with $\vx$, $\vq[{\bm x},z]$; here, we always work with linear theory for which these complications can be ignored (actually, it should be $\vq[{\bm x},z_{\rm dec}]$, where $z_{\rm dec}$ is the redshift at decoupling, but the difference is small and of the same order as other nonlinear terms at recombination that would not be considered anyway). Further, from Eq.~(\ref{eq:theta_r}), one knows that the solution is $\theta_r(\vx, z) = {\theta_{bc,0}(\vx)}/{a}$, where $\theta_{bc,0}$ is the present-day value ($a=1$, $z=0$) of $\theta_r(\vx, z)$. Finally, plugging Eq.~(\ref{eq:delta_r_sol}) and this $\theta_r(\vx, z)$ solution into Eq.~(\ref{eq:other_r}) yields $R_- = \theta_{bc,0}/H_0$ (note that $\dot{D}_r = -H_0/a^2$ and $R_- \propto D_r/\dot{D}_r$, so the $H_0$ normalization in Eq.~(\ref{eq:Dr}) cancels). 

Putting it all together, the relative density and velocity divergence between baryons and CDM are given by
\bq
\label{eq:final_sol_delta} \delta_r(\vx, z) &=& \delta_{bc}(\vx) + \frac{\theta_{bc, 0}(\vx)}{H_0} D_r(z) , \\
\label{eq:final_sol_theta} \theta_{bc}(\vx, z) \equiv \theta_r(\vx, z) &=& \frac{\theta_{bc,0}(\vx)}{a}.
\eq
In Eq.~(\ref{eq:final_sol_theta}), we have introduced the notation $\theta_r \equiv \theta_{bc}$ that is used in the main text (and across most literature). These two equations make apparent that there are two baryon-CDM perturbation modes that can impact galaxy formation: the constant compensated baryon-CDM density perturbation that we studied in the main text, $\delta_{bc}(\vx)$, and the baryon-CDM velocity divergence $\theta_{bc}(\vx, z)$. The latter decays as $a^{-1}$ and its contribution to the $\delta_r$ perturbation becomes also smaller with time because the function $D_r(z)$ is decaying.
{Note that the results derived here are valid in general long after baryon-photon decoupling, regardless of what physical effect (primordial or pre-decoupling) set the initial conditions for $\d_{bc}$ and $\theta_{bc}$.}

\subsection{The power spectra of baryon-CDM perturbations}\label{app:baryon-CDM-theory_2}

{We now focus on baryon-CDM perturbations generated by baryon-photon interactions before decoupling.}
Using Eqs.~(\ref{eq:final_sol_delta}) and (\ref{eq:final_sol_theta}), we can write down the Fourier transform of  $\delta_{bc}$ as (we distinguish Fourier- and real-space quantities by their arguments)
\bq\label{eq:FTdelta_bc}
\delta_{bc}(\vk) = \delta_r(\vk, z) - \frac{aD_r(z)}{H_0}\theta_{bc}(\vk, z).
\eq
Using that $\delta_s(k, z) = (2k^2/(5\Omega_mH_0^2))\R(k)T_s(k, z), s \in \{b, c, m\}$, we have 
\bq\label{eq:delta_r_FT}
\delta_r(\vk, z) = \frac{T_b(k, z) - T_c(k, z)}{T_m(k,z)} \delta_m(\vk, z).
\eq
Further, the velocity divergence term can be worked out as
\bq\label{eq:theta_r_FT}
\theta_{bc}(\vk, z) = i \vk\cdot\vv_{bc} = k\frac{T_{v_{bc}}(k,z)}{T_m(k, z)} \delta_m(\vk, z),
\eq
where the second equality defines the transfer function $T_{v_{bc}}(k,z)$ of the velocity difference between baryons and CDM $\vv_{bc} = \vv_b - \vv_c$. We evaluate all of the transfer functions with the {\sc CAMB} code \cite{camb}. The cross power spectrum between $\delta_m$ and $\delta_{bc}$, $\langle\delta_m(\vk, z)\delta_{bc}({\vk'})\rangle = (2\pi)^3P_{\delta_m\delta_{bc}}(k, z)\delta_D(\vk + \vk')$ is then given by
\bq\label{Pmbc}
P_{\delta_m\delta_{bc}}(k, z) = \left[\frac{T_b(k, z) - T_c(k, z)}{T_m(k,z)} - \frac{kD_r(z)a}{H_0}\frac{T_{v_{bc}}(k, z)}{T_m(k,z)}\right] P_{\delta_m\delta_m}(k, z).
\eq
At the low redshifts we have considered in this paper, the term $\propto T_{v_{bc}}(k,z)$ is only a small contribution ($\approx 4\%$ at $z=3$ and $\approx 2\%$ at $z = 1$, for $k = 0.1\ h/{\rm Mpc}$). Hence, it is a good approximation to discuss the importance of $P_{\delta_m\delta_{bc}}(\vk, z)$ using only the first term (as we did in the discussion below Eq.~(\ref{eq:ratio_impact})); numerically, however, we evaluate $P_{\delta_m\delta_{bc}}$ using all terms in Eq.~(\ref{Pmbc}).  As a side remark, we note that according to Eqs.~(21) and (26) of Ref.~\cite{2017MNRAS.470.2723B}, their $b_{\delta}^{bc}$ parameter multiplies only the $(T_b-T_c)/T_m$ contribution in Eq.~(\ref{Pmbc}). Strictly speaking, the constraints quoted in Ref.~\cite{2017MNRAS.470.2723B} do not correspond to the exact same bias parameter definition, but this does not have any practical consequence given the unimportance of the $T_{v_{bc}}$ term at late times.

The auto power spectrum of $\delta_{bc}$, $P_{\delta_{bc}\delta_{bc}} (k, z)$, can be calculated analogously by evaluating $\langle\delta_{bc}(\vk)\delta_{bc}({\vk'})\rangle$. Noting that $\delta_{m}$ and $\delta_{bc}$ are fully correlated, $P_{\delta_{bc}\delta_{bc}}$ can also be obtained from $P_{\delta_m\delta_{bc}} = \sqrt{P_{\delta_m\delta_m} P_{\delta_{bc}\delta_{bc}}}$, which shows that $P_{\delta_{bc}\delta_{bc}}$ is an even smaller contribution to the galaxy power spectrum than $P_{\delta_{m}\delta_{bc}}$.

\subsection{The generation of $\delta_{bc}$ modes}\label{app:baryon-CDM-theory_3}

\begin{figure}[t!]
        \centering
        \includegraphics[scale = 0.5]{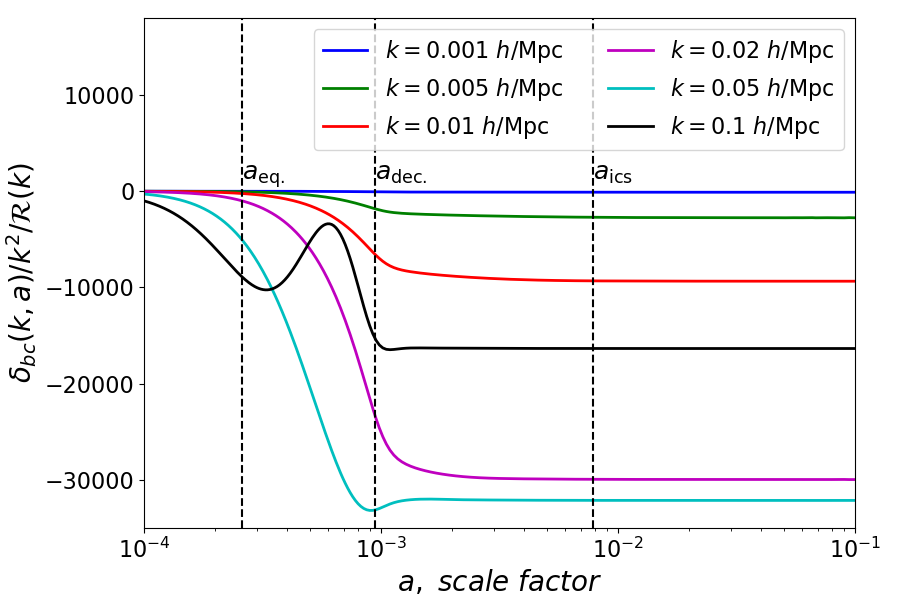}
        \caption{{Time-evolution of $\delta_{bc}(k)$ defined in Eq.~(\ref{eq:FTdelta_bc}) for adiabatic initial conditions after inflation and for different values of $k$, as labeled. The three vertical dashed lines indicate, from left to right, the epoch of matter-radiation equality, photon-baryon decoupling and the starting time of our simulations. The result is obtained with the {\sc CAMB} code and without including the effects of reionization (the latter will induce a time-evolution at $a \gtrsim 0.1$).}}
\label{fig:deltabc}
\end{figure}

{Equations~(\ref{eq:delta_s}) and (\ref{eq:theta_s}) admit a constant mode solution $\delta_{bc}(\vx)$ because they ignore the pressure forces that the baryons feel due to their coupling to the photons, i.e., they are valid only sufficiently after the epoch of photon-baryon decoupling. Prior to this epoch, these forces are sizeable and are what is in fact responsible for generating non-zero $\delta_{bc}$ for adiabatic perturbations after inflation. Figure \ref{fig:deltabc} shows the time evolution of $\delta_{bc}$ as defined in Eq.~(\ref{eq:FTdelta_bc}), but with the coupling between baryons and photons appropriately taken into account using the {\sc CAMB} code. The figure shows that, as expected, the modes that enter the horizon earlier (higher $k$), begin evolving earlier than larger-scale (lower $k$) modes. For example, the $k = 0.1\ h/{\rm Mpc}$ mode has a wavenumber that is larger than the inverse sound horizon at decoupling, so it undergoes more than one full oscillation. Further, sufficiently large-scale modes, specifically modes that are still super-sound horizon by the time of decoupling, display negligible evolution in comparison (cf.~blue line, $k = 0.001\ h/{\rm Mpc}$). Importantly, however, after decoupling, the pressure that drives the generation of $\delta_{bc}$ becomes negligible and all $\delta_{bc}$ modes approach constant values.}

{For the case of baryon-CDM density perturbations generated during inflation, the picture remains qualitatively the same, except that the initial conditions after inflation are not necessarily zero. Specifically, a $\delta_{bc}(k)$ mode with $k \lesssim 0.001\ h/{\rm Mpc}$ generated during inflation will retain approximately the same amplitude at all times, much in the same way as the $k = 0.001\ h/{\rm Mpc}$ mode in Fig.~\ref{fig:deltabc} remains small throughout. It is this case that our separate universe setup strictly applies to, since we generated the initial conditions for the simulations assuming constant modified $\Omega_b$ and $\Omega_c$ at all times up to the starting redshift of the simulation (corresponding to $a_{\rm ics}$ in the figure). The reason why this overestimates the impact on the amplitude of the initial power spectrum of the simulations for $\delta_{bc}$ modes generated by baryon-photon interactions is because these are still growing in between $a_{\rm eq.}$ and $a_{\rm dec.}$, which is when the modified relative abundances of baryons and CDM modify the amplitude of the power spectrum.}

\section{Stellar mass resolution correction factors}\label{app:stecorrection}

\begin{figure}[t!]
        \centering
        \includegraphics[width=\textwidth]{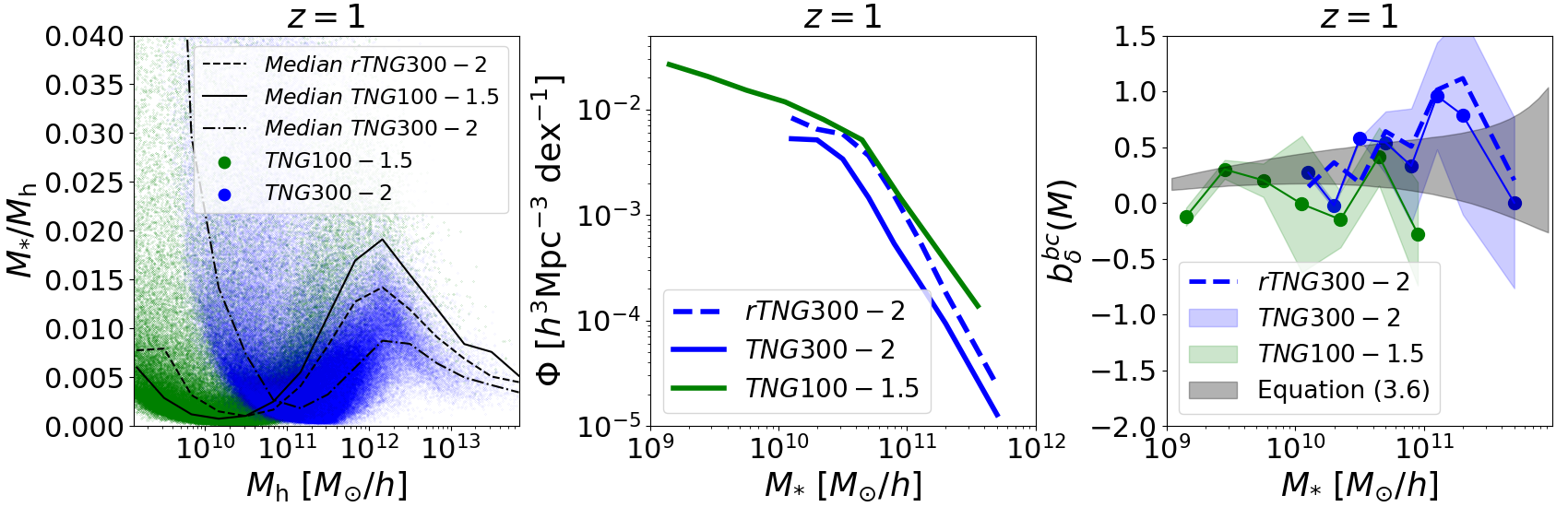}
        \caption{The left panel shows the stellar mass to total halo mass ratios as a function of total halo mass for the TNG100-1.5 and TNG300-2 galaxies. The colored dots indicate the result for all of the galaxies and the black lines indicate the corresponding median relation. The median relation of the resolution-corrected rTNG300-2 galaxies is also shown (dashed line) for comparison with the uncorrected TNG300-2 one (dot-dashed). The middle panel shows the stellar mass function; the dashed blue line shows the result for the rTNG300-2 galaxies, which is in closer agreement (than TNG300-2, solid blue) with the higher-resolution TNG100-1.5 result. The right panel shows $b_{\delta}^{bc}(M_*)$; this is the same as in Fig.~\ref{fig:bcb_stemass_hydro_uncorrected}, but with the result for the corrected rTNG300-2 stellar masses shown as the dashed blue line. All these results are for the \fidu\ cosmology at $z=1$.}
\label{fig:rtng300}
\end{figure}

In this appendix, we describe a stellar mass resolution correction scheme that we apply to the galaxies in our TNG300-2 simulations to test the robustness of our $b_{\delta}^{bc} (M_{*})$ measurements against numerical resolution. The numerical results we show here are for $z=1$, but they are representative of the other redshift values analysed in the paper.

The left panel of Figure \ref{fig:rtng300} shows the ratio $M_{*}/M_{\rm h}$ as a function of $M_{\rm h}$ for the Fiducial cosmology at TNG100-1.5 (green) and TNG300-2 (blue) resolutions at $z=1$. The solid and dot-dashed lines indicate the corresponding median relation in bins of total halo mass. The observed difference between the TNG100-1.5 and TNG300-2 curves reflects the varying levels of convergence of $M_{*}$ at different resolutions. In Ref.~\cite{Pillepich:2017fcc}, the authors demonstrate that a correction factor constructed using the stellar-to-halo-mass relations of two same-volume IllustrisTNG resolutions works well in bringing the results of simulations done at different volumes together in other stellar-mass-related quantities such as the stellar mass function or stellar mass radial profiles (cf.~Appendix A of Ref.~\cite{Pillepich:2017fcc} for more details). Here, we apply the same resolution correction strategy to the TNG300-2 catalogues. Specifically, we define a stellar mass correction factor as
\bq
\label{eq:corr2}C^\text{TNG300-2}\left(M_{\rm h}\right) &=& \frac{\langle M_{*}^\text{oTNG100-2}(M_{\rm h})\rangle_{\rm median}}{\langle M_{*}^\text{oTNG100-3}(M_{\rm h})\rangle_{\rm median}},
\eq
where the numerator and denominator on the right-hand side are the median relations in the original TNG100-2 and TNG100-3 simulations, respectively; the superscripts ${}^\text{oTNG100-2}$ and ${}^\text{oTNG100-3}$ stress that we use the original simulations, which have matching phases of the initial conditions. It is important to note also that the TNG300-2 ($N_p = 1250^3$, $L_{\rm box} = 205\ {\rm Mpc/h}$) and TNG100-3 ($N_p = 455^3$, $L_{\rm box} = 75\ {\rm Mpc/h}$) simulations have approximately the same mass resolution. Our corrected stellar mass catalogues are then subsequently obtained by multiplying all of the stellar mass values of the TNG300-2 galaxies by $C^\text{TNG300-2}\left(M_{\rm h}\right)$; specifically, we interpolate over the values of $M_{\rm h}$ in Eq.~(\ref{eq:corr2}), which is defined only at a finite number of total mass bins. We label the results of the corrected TNG300-2 catalogues as rTNG300-2, whose stellar mass values should be representative of a TNG100-2 resolution. While this still falls short of our higher resolution TNG100-1.5 results, it nonetheless allows us to check whether numerical convergence at the level of stellar masses plays a critical role in our measurements.

The dashed black line in the left panel of Fig.~\ref{fig:rtng300} shows the resulting median relation of the rTNG300-2 galaxy catalogues. Further, the middle panel of Fig.~\ref{fig:rtng300} shows the TNG300-2 stellar mass function measured with (dashed blue) and without (solid blue) the stellar mass correction.  As expected, for both the stellar-to-halo-mass relation and the stellar mass function,  the rTNG300-2 results are in closer agreement with TNG100-1.5, compared to TNG300-2. More importantly, the right panel of Fig.~\ref{fig:rtng300} shows the same as the $z=1$ panel of Fig.~\ref{fig:bcb_stemass_hydro_uncorrected}, but with the result of the corrected rTNG300-2 galaxies shown as well (dashed line). The $b_{\delta}^{bc}(M_*)$ measured from the TNG300-2 and rTNG300-2 catalogues display only small differences with one another; in particular, both agree well with the higher-resolution TNG100-1.5 result. The discussion in the main body of the paper (in Sec.~\ref{sec:res:stemass}) about the TNG300-2 catalogues thus holds equally to the case of the corrected rTNG300-2 ones. This is not surprising since we have corrected the ${\rm \fidu}$, \high\ and \loww\ cosmologies using the same correction factors, which preserve the relative difference between cosmologies that is effectively measured by $b_{\delta}^{bc}$; the observed small differences are caused by galaxies moving to different stellar mass bins.

\bibliography{REFS}

\def\eprinttmppp@#1arXiv:@{#1}
\providecommand{\arxivlink[1]}{\href{http://arxiv.org/abs/#1}{arXiv:#1}}
\providecommand{\arxivlinknopre[1]}{\href{http://arxiv.org/abs/#1}{#1}}
\providecommand{\eprintmod}[1][XXXX.XXXX]{\IfSubStr{#1}{arXiv}{\arxivlinknopre{#1}}{\arxivlink{#1}}}
\providecommand{\adsurl}[1]{\href{#1}{ADS}}
\begin{thebibliography}{89}
\expandafter\ifx\csname natexlab\endcsname\relax\def\natexlab#1{#1}\fi
\expandafter\ifx\csname bibnamefont\endcsname\relax
  \def\bibnamefont#1{#1}\fi
\expandafter\ifx\csname bibfnamefont\endcsname\relax
  \def\bibfnamefont#1{#1}\fi
\expandafter\ifx\csname citenamefont\endcsname\relax
  \def\citenamefont#1{#1}\fi
\expandafter\ifx\csname url\endcsname\relax
  \def\url#1{\texttt{#1}}\fi
\expandafter\ifx\csname urlprefix\endcsname\relax\def\urlprefix{URL }\fi

\bibitem{biasreview}
V.~{Desjacques}, D.~{Jeong} and F.~{Schmidt},
\newblock \physrep {\bf 733}, 1 (2018), [\eprintmod[1611.09787]].

\bibitem{2018JCAP...12..035D}
V.~{Desjacques}, D.~{Jeong} and F.~{Schmidt},
\newblock Journal of Cosmology and Astro-Particle Physics {\bf 2018}, 035
  (2018), [\eprintmod[1806.04015]].

\bibitem{fry/gaztanaga:1983}
J.~N. {Fry} and E.~{Gaztanaga},
\newblock \apj {\bf 413}, 447 (1993), [\eprintmod[arXiv:astro-ph/9302009]].

\bibitem{Bernardeau/etal:2002}
F.~{Bernardeau}, S.~{Colombi}, E.~{Gazta{\~n}aga} and R.~{Scoccimarro},
\newblock \physrep {\bf 367}, 1 (2002), [\eprintmod[arXiv:astro-ph/0112551]].

\bibitem{mcdonald/roy:2009}
P.~{McDonald} and A.~{Roy},
\newblock \jcap {\bf 8}, 20 (2009), [\eprintmod[0902.0991]].

\bibitem{chan/scoccimarro/sheth:2012}
K.~C. {Chan}, R.~{Scoccimarro} and R.~K. {Sheth},
\newblock \prd {\bf 85}, 083509 (2012), [\eprintmod[1201.3614]].

\bibitem{baldauf/etal:2012}
T.~{Baldauf}, U.~{Seljak}, V.~{Desjacques} and P.~{McDonald},
\newblock ArXiv e-prints  (2012), [\eprintmod[1201.4827]].

\bibitem{saito/etal:14}
S.~{Saito} {\em et~al.},
\newblock \prd {\bf 90}, 123522 (2014), [\eprintmod[1405.1447]].

\bibitem{2019arXiv190411294L}
T.~{Lazeyras} and F.~{Schmidt},
\newblock arXiv e-prints , arXiv:1904.11294 (2019), [\eprintmod[1904.11294]].

\bibitem{grin/dore/kamionkowski}
D.~{Grin}, O.~{Dor{\'e}} and M.~{Kamionkowski},
\newblock \prd {\bf 84}, 123003 (2011), [\eprintmod[1107.5047]].

\bibitem{1997PhRvD..56..535L}
A.~{Linde} and V.~{Mukhanov},
\newblock \prd {\bf 56}, R535 (1997), [\eprintmod[astro-ph/9610219]].

\bibitem{2000PhRvD..62d3504L}
D.~{Langlois} and A.~{Riazuelo},
\newblock \prd {\bf 62}, 043504 (2000), [\eprintmod[astro-ph/9912497]].

\bibitem{2003PhRvD..67b3503L}
D.~H. {Lyth}, C.~{Ungarelli} and D.~{Wands},
\newblock \prd {\bf 67}, 023503 (2003), [\eprintmod[astro-ph/0208055]].

\bibitem{2006RvMP...78..537B}
B.~A. {Bassett}, S.~{Tsujikawa} and D.~{Wands},
\newblock Reviews of Modern Physics {\bf 78}, 537 (2006),
  [\eprintmod[astro-ph/0507632]].

\bibitem{barkana/loeb:11}
R.~{Barkana} and A.~{Loeb},
\newblock \mnras {\bf 415}, 3113 (2011), [\eprintmod[1009.1393]].

\bibitem{2016PhRvD..94f3508S}
F.~{Schmidt},
\newblock \prd {\bf 94}, 063508 (2016), [\eprintmod[1602.09059]].

\bibitem{2016ApJ...830...68A}
K.~{Ahn},
\newblock \apj {\bf 830}, 68 (2016), [\eprintmod[1603.09356]].

\bibitem{2019JCAP...06..006C}
S.-F. {Chen}, E.~{Castorina} and M.~{White},
\newblock \jcap {\bf 2019}, 006 (2019), [\eprintmod[1903.00437]].

\bibitem{tseliakhovich/hirata:2010}
D.~{Tseliakhovich} and C.~{Hirata},
\newblock \prd {\bf 82}, 083520 (2010), [\eprintmod[1005.2416]].

\bibitem{blazek/etal:15}
J.~A. {Blazek}, J.~E. {McEwen} and C.~M. {Hirata},
\newblock Physical Review Letters {\bf 116}, 121303 (2016),
  [\eprintmod[1510.03554]].

\bibitem{2010MNRAS.401..791S}
V.~{Springel},
\newblock \mnras {\bf 401}, 791 (2010), [\eprintmod[0901.4107]].

\bibitem{2016MNRAS.455.1134P}
R.~{Pakmor} {\em et~al.},
\newblock \mnras {\bf 455}, 1134 (2016), [\eprintmod[1503.00562]].

\bibitem{2017MNRAS.465.3291W}
R.~{Weinberger} {\em et~al.},
\newblock \mnras {\bf 465}, 3291 (2017), [\eprintmod[1607.03486]].

\bibitem{Pillepich:2017jle}
A.~Pillepich {\em et~al.},
\newblock Mon. Not. Roy. Astron. Soc. {\bf 473}, 4077 (2018),
  [\eprintmod[1703.02970]].

\bibitem{2017MNRAS.470.2723B}
F.~{Beutler}, U.~{Seljak} and Z.~{Vlah},
\newblock \mnras {\bf 470}, 2723 (2017), [\eprintmod[1612.04720]].

\bibitem{soumagnac/etal:16}
M.~T. {Soumagnac} {\em et~al.},
\newblock PRL {\bf 116}, 201302 (2016), [\eprintmod[1602.01839]].

\bibitem{2019MNRAS.485.1248S}
M.~T. {Soumagnac}, C.~G. {Sabiu}, R.~{Barkana} and J.~{Yoo},
\newblock \mnras {\bf 485}, 1248 (2019), [\eprintmod[1802.10368]].

\bibitem{yoo/etal:2011}
J.~{Yoo}, N.~{Dalal} and U.~{Seljak},
\newblock \jcap {\bf 7}, 18 (2011), [\eprintmod[1105.3732]].

\bibitem{yoo/seljak}
J.~{Yoo} and U.~{Seljak},
\newblock \prd {\bf 88}, 103520 (2013), [\eprintmod[1308.1401]].

\bibitem{tseliakhovich/barkana/hirata}
D.~{Tseliakhovich}, R.~{Barkana} and C.~M. {Hirata},
\newblock \mnras {\bf 418}, 906 (2011), [\eprintmod[1012.2574]].

\bibitem{dalal/etal:2010}
N.~{Dalal}, U.-L. {Pen} and U.~{Seljak},
\newblock \jcap {\bf 11}, 7 (2010), [\eprintmod[1009.4704]].

\bibitem{slepian/eisenstein}
Z.~{Slepian} and D.~J. {Eisenstein},
\newblock \mnras {\bf 448}, 9 (2015), [\eprintmod[1411.4052]].

\bibitem{asaba/ichiki/tashiro}
S.~{Asaba}, K.~{Ichiki} and H.~{Tashiro},
\newblock \prd {\bf 93}, 023518 (2016), [\eprintmod[1508.07719]].

\bibitem{Slepian:2016nfb}
Z.~Slepian {\em et~al.},
\newblock MNRAS {\bf 474}, 2109 (2018), [\eprintmod[1607.06098]].

\bibitem{popa/etal}
C.~{Popa}, S.~{Naoz}, F.~{Marinacci} and M.~{Vogelsberger},
\newblock ArXiv e-prints  (2015), [\eprintmod[1512.06862]].

\bibitem{2011ApJ...736..147G}
T.~H. {Greif}, S.~D.~M. {White}, R.~S. {Klessen} and V.~{Springel},
\newblock \apj {\bf 736}, 147 (2011), [\eprintmod[1101.5493]].

\bibitem{2011MNRAS.412L..40M}
U.~{Maio}, L.~V.~E. {Koopmans} and B.~{Ciardi},
\newblock \mnras {\bf 412}, L40 (2011), [\eprintmod[1011.4006]].

\bibitem{2011ApJ...730L...1S}
A.~{Stacy}, V.~{Bromm} and A.~{Loeb},
\newblock \apj {\bf 730}, L1 (2011), [\eprintmod[1011.4512]].

\bibitem{2012ApJ...747..128N}
S.~{Naoz}, N.~{Yoshida} and N.~Y. {Gnedin},
\newblock \apj {\bf 747}, 128 (2012), [\eprintmod[1108.5176]].

\bibitem{Visbal/etal:12}
E.~{Visbal}, R.~{Barkana}, A.~{Fialkov}, D.~{Tseliakhovich} and C.~M. {Hirata},
\newblock \nat {\bf 487}, 70 (2012), [\eprintmod[1201.1005]].

\bibitem{2012ApJ...760....4O}
R.~M. {O'Leary} and M.~{McQuinn},
\newblock \apj {\bf 760}, 4 (2012), [\eprintmod[1204.1344]].

\bibitem{2013ApJ...763...27N}
S.~{Naoz}, N.~{Yoshida} and N.~Y. {Gnedin},
\newblock \apj {\bf 763}, 27 (2013), [\eprintmod[1207.5515]].

\bibitem{2013ApJ...771...81R}
M.~L.~A. {Richardson}, E.~{Scannapieco} and R.~J. {Thacker},
\newblock \apj {\bf 771}, 81 (2013), [\eprintmod[1305.3276]].

\bibitem{2019ApJ...878L..23C}
Y.~S. {Chiou}, S.~{Naoz}, B.~{Burkhart}, F.~{Marinacci} and M.~{Vogelsberger},
\newblock \apj {\bf 878}, L23 (2019), [\eprintmod[1904.08941]].

\bibitem{Fialkov:2014rba}
A.~Fialkov,
\newblock Int. J. Mod. Phys. {\bf D23}, 1430017 (2014),
  [\eprintmod[1407.2274]].

\bibitem{shoji/komatsu}
M.~{Shoji} and E.~{Komatsu},
\newblock \apj {\bf 700}, 705 (2009), [\eprintmod[0903.2669]].

\bibitem{somogyi/smith:2010}
G.~{Somogyi} and R.~E. {Smith},
\newblock \prd {\bf 81}, 023524 (2010), [\eprintmod[0910.5220]].

\bibitem{bernardeau/vdr/vernizzi}
F.~{Bernardeau}, N.~{Van de Rijt} and F.~{Vernizzi},
\newblock \prd {\bf 87}, 043530 (2013), [\eprintmod[1209.3662]].

\bibitem{lewandowski/perko/senatore}
M.~{Lewandowski}, A.~{Perko} and L.~{Senatore},
\newblock \jcap {\bf 5}, 019 (2015), [\eprintmod[1412.5049]].

\bibitem{2009PhRvD..80f3535G}
C.~{Gordon} and J.~R. {Pritchard},
\newblock \prd {\bf 80}, 063535 (2009), [\eprintmod[0907.5400]].

\bibitem{2010ApJ...716..907H}
G.~P. {Holder}, K.~M. {Nollett} and A.~e. {van Engelen},
\newblock \apj {\bf 716}, 907 (2010), [\eprintmod[0907.3919]].

\bibitem{2014PhRvD..89b3006G}
D.~{Grin}, D.~{Hanson}, G.~P. {Holder}, O.~{Dor{\'e}} and M.~{Kamionkowski},
\newblock \prd {\bf 89}, 023006 (2014), [\eprintmod[1306.4319]].

\bibitem{2017PhRvD..96h3508S}
T.~L. {Smith}, J.~B. {Mu{\~n}oz}, R.~{Smith}, K.~{Yee} and D.~{Grin},
\newblock \prd {\bf 96}, 083508 (2017), [\eprintmod[1704.03461]].

\bibitem{2016PhRvD..93d3008M}
J.~B. {Mu{\~n}oz}, D.~{Grin}, L.~{Dai}, M.~{Kamionkowski} and E.~D. {Kovetz},
\newblock \prd {\bf 93}, 043008 (2016), [\eprintmod[1511.04441]].

\bibitem{2019arXiv190400024H}
C.~{Heinrich} and M.~{Schmittfull},
\newblock arXiv e-prints , arXiv:1904.00024 (2019), [\eprintmod[1904.00024]].

\bibitem{2019JCAP...05..031C}
G.~{Cabass} and F.~{Schmidt},
\newblock Journal of Cosmology and Astro-Particle Physics {\bf 2019}, 031
  (2019), [\eprintmod[1812.02731]].

\bibitem{li/hu/takada}
Y.~{Li}, W.~{Hu} and M.~{Takada},
\newblock \prd {\bf 89}, 083519 (2014), [\eprintmod[1401.0385]].

\bibitem{2014PhRvD..90j3530L}
Y.~{Li}, W.~{Hu} and M.~{Takada},
\newblock \prd {\bf 90}, 103530 (2014), [\eprintmod[1408.1081]].

\bibitem{wagner/etal:2014}
C.~Wagner, F.~Schmidt, C.-T. Chiang and E.~Komatsu,
\newblock Mon. Not. Roy. Astron. Soc. {\bf 448}, L11 (2015),
  [\eprintmod[1409.6294]].

\bibitem{CFCpaper2}
L.~{Dai}, E.~{Pajer} and F.~{Schmidt},
\newblock \jcap {\bf 10}, 059 (2015), [\eprintmod[1504.00351]].

\bibitem{baldauf/etal:2015}
T.~{Baldauf}, U.~{Seljak}, L.~{Senatore} and M.~{Zaldarriaga},
\newblock \jcap {\bf 9}, 007 (2016), [\eprintmod[1511.01465]].

\bibitem{response}
C.~Wagner, F.~Schmidt, C.-T. Chiang and E.~Komatsu,
\newblock JCAP {\bf 1508}, 042 (2015), [\eprintmod[1503.03487]].

\bibitem{lazeyras/etal}
T.~{Lazeyras}, C.~{Wagner}, T.~{Baldauf} and F.~{Schmidt},
\newblock \jcap {\bf 2}, 018 (2016), [\eprintmod[1511.01096]].

\bibitem{li/hu/takada:2016}
Y.~{Li}, W.~{Hu} and M.~{Takada},
\newblock \prd {\bf 93}, 063507 (2016), [\eprintmod[1511.01454]].

\bibitem{2018PhRvD..97l3526C}
C.-T. {Chiang}, W.~{Hu}, Y.~{Li} and M.~{LoVerde},
\newblock \prd {\bf 97}, 123526 (2018), [\eprintmod[1710.01310]].

\bibitem{2019arXiv190402070B}
A.~{Barreira} {\em et~al.},
\newblock arXiv e-prints , arXiv:1904.02070 (2019), [\eprintmod[1904.02070]].

\bibitem{2014MNRAS.445..175G}
S.~{Genel} {\em et~al.},
\newblock \mnras {\bf 445}, 175 (2014), [\eprintmod[1405.3749]].

\bibitem{2014MNRAS.444.1518V}
M.~{Vogelsberger} {\em et~al.},
\newblock \mnras {\bf 444}, 1518 (2014), [\eprintmod[1405.2921]].

\bibitem{2018MNRAS.480.5113M}
F.~{Marinacci} {\em et~al.},
\newblock \mnras {\bf 480}, 5113 (2018), [\eprintmod[1707.03396]].

\bibitem{Pillepich:2017fcc}
A.~Pillepich {\em et~al.},
\newblock Mon. Not. Roy. Astron. Soc. {\bf 475}, 648 (2018),
  [\eprintmod[1707.03406]].

\bibitem{2018MNRAS.477.1206N}
J.~P. {Naiman} {\em et~al.},
\newblock \mnras {\bf 477}, 1206 (2018), [\eprintmod[1707.03401]].

\bibitem{2018MNRAS.475..676S}
V.~{Springel} {\em et~al.},
\newblock \mnras {\bf 475}, 676 (2018), [\eprintmod[1707.03397]].

\bibitem{Nelson:2017cxy}
D.~Nelson {\em et~al.},
\newblock Mon. Not. Roy. Astron. Soc. {\bf 475}, 624 (2018),
  [\eprintmod[1707.03395]].

\bibitem{Nelson:2018uso}
D.~Nelson {\em et~al.},
\newblock arXiv:1812.05609  (2018), [\eprintmod[1812.05609]].

\bibitem{2015ascl.soft02003S}
V.~{Springel},
\newblock {N-GenIC: Cosmological structure initial conditions},
\newblock Astrophysics Source Code Library, 2015, \eprintmod[1502.003].

\bibitem{camb}
A.~{Lewis}, A.~{Challinor} and A.~{Lasenby},
\newblock \apj {\bf 538}, 473 (2000), [\eprintmod[astro-ph/9911177]].

\bibitem{2003MNRAS.344..481Y}
N.~{Yoshida}, N.~{Sugiyama} and L.~{Hernquist},
\newblock \mnras {\bf 344}, 481 (2003), [\eprintmod[astro-ph/0305210]].

\bibitem{2013MNRAS.434.1756A}
R.~E. {Angulo}, O.~{Hahn} and T.~{Abel},
\newblock \mnras {\bf 434}, 1756 (2013), [\eprintmod[1301.7426]].

\bibitem{2017MNRAS.467.4401V}
W.~{Valkenburg} and F.~{Villaescusa-Navarro},
\newblock \mnras {\bf 467}, 4401 (2017), [\eprintmod[1610.08501]].

\bibitem{2001MNRAS.328..726S}
V.~{Springel}, S.~D.~M. {White}, G.~{Tormen} and G.~{Kauffmann},
\newblock \mnras {\bf 328}, 726 (2001), [\eprintmod[astro-ph/0012055]].

\bibitem{2008ApJ...688..709T}
J.~{Tinker} {\em et~al.},
\newblock \apj {\bf 688}, 709 (2008), [\eprintmod[0803.2706]].

\bibitem{2016MNRAS.460.1457S}
S.~{Saito} {\em et~al.},
\newblock \mnras {\bf 460}, 1457 (2016), [\eprintmod[1509.00482]].

\bibitem{2010ApJ...724..878T}
J.~L. {Tinker} {\em et~al.},
\newblock \apj {\bf 724}, 878 (2010), [\eprintmod[1001.3162]].

\bibitem{2017MNRAS.466.2242B}
F.~{Beutler} {\em et~al.},
\newblock \mnras {\bf 466}, 2242 (2017), [\eprintmod[1607.03150]].

\bibitem{2013arXiv1308.0847L}
M.~{Levi} {\em et~al.},
\newblock ArXiv e-prints  (2013), [\eprintmod[1308.0847]].

\bibitem{2019arXiv190808953H}
S.~C. {Hotinli}, J.~B. {Mertens}, M.~C. {Johnson} and M.~{Kamionkowski},
\newblock arXiv e-prints , arXiv:1908.08953 (2019), [\eprintmod[1908.08953]].

\bibitem{2001PhRvD..63f3002K}
E.~{Komatsu} and D.~N. {Spergel},
\newblock \prd {\bf 63}, 063002 (2001), [\eprintmod[arXiv:astro-ph/0005036]].

\bibitem{2014arXiv1412.4872D}
O.~{Dor{\'e}} {\em et~al.},
\newblock arXiv e-prints , arXiv:1412.4872 (2014), [\eprintmod[1412.4872]].

\end{thebibliography}

\end{document}